\documentclass[showpacs,preview,preprintnumbers,amsmath,amssymb]{revtex4}

\usepackage{graphics}
\usepackage{graphicx}
\usepackage{epsfig}
\usepackage{epsf}
\usepackage{bm}
\usepackage{subfigure}
\usepackage{amsmath}
\usepackage{color}

\usepackage{amssymb}

\begin{document}




\title{Laser Pulsing in Linear Compton Scattering}



\author{G.~A.~Krafft$^{1,2}$\thanks{Email: krafft@jlab.org}, E.~Johnson$^{1}$,
K.~Deitrick$^{1}$, B.~Terzi{\'c}$^{1}$,
R.~Kelmar$^{3}$, T.~Hodges$^{4}$, W.~Melnitchouk$^{2}$, J.~R.~Delayen$^{1,2}$}

\affiliation{
$^1$Department of Physics, Center for Accelerator Science, Old Dominion University,
Norfolk, Virginia 23529 \\
$^2$Thomas Jefferson National Accelerator Facility, Newport News, Virginia 23606 \\
$^3$Department of Physics, Union College, Schenectady, New York 12308 \\
$^4$Department of Physics, Arizona State University, Tempe, Arizona 85004}


\begin{abstract}
Previous work on calculating energy spectra from Compton scattering events has either neglected considering the pulsed structure of the incident laser beam, or has calculated these effects in an approximate way subject to criticism. In this paper, this problem has been reconsidered within a linear plane wave model for the incident laser beam. By performing the proper Lorentz transformation of the Klein-Nishina scattering cross section, a spectrum calculation can be created which allows the electron beam energy spread and emittance effects on the spectrum to be accurately calculated, essentially by summing over the emission of each individual electron. Such an approach has the obvious advantage that it is easily integrated with a particle distribution generated by particle tracking, allowing precise calculations of spectra for realistic particle distributions ``in collision". The method is used to predict the energy spectrum of radiation passing through an aperture for the proposed Old Dominion University inverse Compton source. Many of the results allow easy scaling estimates to be made of the expected spectrum.

\end{abstract}


\maketitle

\section{Introduction} \label{sec:intro}

Compton or Thomson scattering can be used in constructing sources of high energy photons \cite{krafftreview, ruth, p68, betal72}.
In recent years there has been a revival of activity in the subject driven by the desire to produce several keV X-ray sources from relatively compact relativistic electron accelerators.
Such sources are attractive due to the narrow bandwidth generated in the output radiation.
A group at Old Dominion University (ODU) and Jefferson Lab has been actively engaged in designing such a source \cite{sato,IPAC2015}.
As part of the design process, it is important to quantify the effect of electron beam energy spread, electron beam emittance, and the finite laser pulse length on the radiation generated. In the course of our design process we have developed a calculation method yielding the energy spectral distribution of the radiation produced by the scattering event, and extended it so that the radiation from a bunch of relativistic electrons may be obtained. In this paper we summarize the calculation method, and use it in a benchmarking calculation to confirm several results previously published \cite{slrtw09}. In addition, we use the method to suggest a needed correction in Ref.~\cite{gheb}, and to make predictions regarding X-ray source performance for a compact Superconducting RF (SRF) linac based source proposed at ODU. The calculations show that the expected brilliance from this source will be world-leading for Compton sources.

Our calculation method is somewhat different from others \cite{hb12,hw13} because the incident laser electromagnetic field is specified as an input to the calculation through the normalized vector potential. Thus the finite pulse effects possible in a real laser pulse will be modeled properly within a plane-wave approximation. The flat-pulse approximation is not adopted \cite{esary}, although this case can be encompassed within the method. Likewise, it is not necessary to characterize the incident photon beam only by a series of moments. More flexibility is allowed through investigating various models for the vector potential. The approach in this calculation is closest to that of Petrillo {\it et al.} \cite{ELI1}. We note, however, that some modifications of their published calculations are needed. On the other hand, we confirm their results, with some exceptions noted, by also calculating with parameters for the Extreme Light Infrastructure (ELI) - Nuclear Physics \cite{ELI2}.

We report on calculations completed using the quantum mechanical Klein-Nishina \cite{klni} cross section (higher order quantum effects are neglected) under the assumption that the incident laser field is a plane wave. The number density of incident photons is related to the wave function of the incident field using the usual semiclassical approach. As such, this approximation is invalid in situations with highest field
strengths where multiphoton quantum emission can occur. In contrast to our previous work on this subject \cite{terzic} the full Compton recoil is included, from which the linear Thomson scattering results are recovered properly.

As a result of our design work, a literature search concerning the scattering of circularly polarized laser beams was undertaken. Perhaps surprisingly, although the case of linear polarization is extremely well documented in text books \cite{bd,sak,itzzub}, the case of scattering circularly or elliptically polarized beams is not so well documented. More disturbingly, there are misleading and/or incorrect solutions to this problem given in fairly well-known references. In this paper a proper solution to the problem of the Compton scattering of circularly polarized light is presented in a reasonably convenient general form. Our results are consistent with the recent discussion in Ref.~\cite{bocaf1}.

The paper is organized as follows. In Section \ref{sec:spectra1}, the spectral distribution of interest is defined and a single electron emission spectrum is derived for the full Compton effect.
Next, in Section \ref{sec:nummeth}, the method used to numerically integrate the individual electron spectra, and the method to add up and average the emission from a compressed bunch of electrons, are given. The main body of numerical results is presented in Sections \ref{sec:bmv} and \ref{sec:LPeffects}. Here, a series of benchmarking studies and results are recorded, and a generalization of a scaling law discussed by several authors \cite{ELI1,slrtw09,krafftreview,haj1} is given and verified numerically. In a previous publication \cite{gheb}, a calculation using the Thomson limit was documented. In Section \ref{sec:fshifting} this calculation is shown to be more appropriately completed using the full Compton recoil, and the modification of the emission spectrum in this case is documented. In Section \ref{sec:ODUsource} the ODU compact Compton source design is evaluated by taking front-to-end simulation data of the beam produced at the interaction point in the source, and using it to predict the photon spectrum in collision. In the final technical section, Sec.~\ref{sec:circular}, the modifications needed to properly calculate the circularly polarized case are given.  Finally, the importance of the
new results is discussed and conclusions drawn in Section \ref{sec:summary}.

\section{Energy Spectral Distributions for the Full Compton Effect} \label{sec:spectra1}

	Calculations of synchrotron radiation from various arrangements of magnets has an extensive literature.
The result of Co\"isson on the energy spectral distribution of synchrotron radiation produced
by an electron traversing a ``short" magnet is a convenient starting point for our calculations \cite{coissoncite}.
In MKS units and translating his expressions into the symbols used in this paper, the result for the spectrum of the energy radiated by a single particle $U_\gamma$ into a given solid angle $\mathrm{d}\Omega$ is (see \cite{krafft} for more
detailed discussion and cgs expressions)
\begin{equation}
\frac{\mathrm{d}^2U_\gamma}{\mathrm{d}\omega' \mathrm{d}\Omega}=\frac{r_e^2\epsilon_0c}{2\pi}
\left|\widetilde{B}[\omega'(1-\beta\cos\theta)/\beta c]\right|^2\frac{(1-\beta\cos\theta)^2\sin^2\phi+(\cos\theta-\beta)^2\cos^2\phi
}{\gamma^2(1-\beta\cos\theta)^4},
\end{equation}
where $\widetilde{B}$ is the spatial Fourier transform of the
transverse magnetic field bending the electron evaluated
in the
lab frame, and the notation indicates the transform is evaluated at the Doppler shifted wave number
$\omega'(1-\beta\cos\theta)/c\beta$.
Here $\epsilon_0$ is the free-space permitivity, $r_e=e^2/(4\pi\epsilon_0mc^2)$ is the classical electron radius ($\approx$ 2.82$\times10^{-15}$ m), $c$ is the velocity of light, $\beta=v_z/c$ the relativistic longitudinal velocity, and $\gamma=1/\sqrt{1-\beta^2}$ is the usual relativistic factor. Following standard treatments \cite{bd,sak,itzzub}, $\omega'$ denotes the scattered photon angular frequency as measured in the lab frame, and $\theta$ and $\phi$ are the standard polar angles of the scattered radiation in a coordinate system whose $z$-axis is aligned with the electron velocity.

A similar expression applies for linear Thomson scattering.
The energy spectral density of the output pulse scattered by an electron may
be computed analytically in the linear Thomson backscatter
limit as
\begin{equation} \label{thomlt}
\frac{\mathrm{d}^2U_\gamma}{\mathrm{d}\omega' \mathrm{d}\Omega}=\frac{r_e^2\epsilon_0}{2\pi c}
\left|\widetilde{E}[\omega'(1-\beta\cos\theta)/c(1+\beta)]\right|^2\frac{(1-\beta\cos\theta)^2\sin^2\phi+(\cos\theta-\beta)^2\cos^2\phi
}{\gamma^2(1-\beta\cos\theta)^4}.
\end{equation}

As will be shown below, in the Thomson limit the electron recoil is neglected in the scattering event. This limit is valid for many X-ray source designs (ours included), but starts to break down in some of the higher electron energy sources being considered \cite{ELI1,gheb}. Thus, in this section, the spectral distribution is calculated including the full Compton recoil for plane wave incident laser pulses. Implicit in the derivations is that linear scattering applies, $a(z)\ll 1$, where $a=eA_x/mc$ is the normalized vector potential for the incident pulse. This assumption will be adopted throughout this paper.

In the calculations of the scattered energy a semiclassical model for the wave function of the incident laser is taken and a plane wave model for this field is adopted. The latter assumption is justified in our work because the collision point source size in our designs is much smaller than the collimation aperture for the X-rays produced: there is relatively little error introduced in replacing the actual scattering angle with the angle to the observation location in the far field limit. In the plane wave approximation the vector potential and electric field of the incident laser pulse are represented as wave packets
\begin{equation}
A_x \left( {z,t} \right) = \frac{1}{{2\pi }}\int_{-\infty }^\infty {\widetilde A_x\left( \omega  \right)} e^{i\omega \left( {z/c + t} \right)} \mathrm{d}\omega ,
\end{equation}

\begin{equation}
E_x \left( {z,t} \right) = -\frac{1}{{2\pi }}\int_{- \infty }^\infty  i\omega {\widetilde A_x\left( \omega  \right)} e^{i\omega \left( {z/c + t} \right)} \mathrm{d}\omega ,
\end{equation}
with
\begin{equation}
\widetilde A_x\left( \omega  \right) = \int_{- \infty }^\infty  {A_x \left( {z = 0,t} \right)} e^{ - i\omega t} \mathrm{d}t.
\end{equation}
The power per unit area in the wave packet is
\begin{equation}
c\left[ {\frac{{\epsilon _0 }}{2}E_x^2 \left( {z,t} \right) + \frac{{B_y^2 \left( {z,t} \right)}}{{2\mu _0 }}} \right] = \epsilon _0 cE_x^2 \left( {z,t} \right).
\end{equation}
Because of Parseval's theorem, the time-integrated intensity or energy per area in the pulse passing by an electron moving along the $z$-axis of the coordinate system is
\begin{equation}
\int_{- \infty }^\infty  {E_x^2 \left( {z = 0,t} \right)} \mathrm{d}t = \frac{1}{2\pi }\int_{- \infty }^\infty  \left|\widetilde E_x\left( {\omega } \right)\right|^2 \mathrm{d}\omega ,
\end{equation}
where $\widetilde E_x(\omega)$ now denotes the Fourier time transform of the incident pulse. The incident energy per unit angular frequency per unit area is thus
\begin{equation}
\frac{{\epsilon _0 c}}{{2\pi }}\left| {\widetilde E_x\left( \omega  \right)} \right|^2 =
\frac{{\epsilon _0 c}}{{2\pi }}\left| {\omega\widetilde A_x\left( \omega  \right)} \right|^2 .
\end{equation}

Within a ``semi-classical" analysis the number of incident photons per unit angular frequency per area is consequently
\begin{equation}
\frac{{\epsilon _0 c}}{{2\pi }}\frac{{\left| {\widetilde E_x\left( \omega  \right)} \right|^2 }}{{\hbar \left| \omega  \right|}}.
\end{equation}
The number of scattered photons generated into a given solid angle $\mathrm{d}\Omega$ is
\begin{equation}
\frac{{\mathrm{d}N_{scat} }}{{\mathrm{d}\Omega }} = \int_{- \infty }^\infty  {\frac{{\epsilon _0 c}}{{2\pi }}\frac{{\left| {\widetilde E_x\left( \omega  \right)} \right|^2 }}{{\hbar \left| \omega  \right|}}\frac{{\mathrm{d}\sigma }}{{\mathrm{d}\Omega }}} \, \mathrm{d}\omega ,
\end{equation}
and, because the scattered photon has energy $\hbar\omega'$, the total scattered energy is
\begin{equation}
\frac{{\mathrm{d}U_\gamma }}{{\mathrm{d}\Omega }} = \int_{ - \infty }^\infty  {\frac{{\epsilon _0 c}}{{2\pi }}\left| {\widetilde E_x\left( \omega  \right)} \right|^2 {\frac{{\omega '}}{\omega }}\frac{{\mathrm{d}\sigma }}{{\mathrm{d}\Omega }}} \, \mathrm{d}\omega
\end{equation}
where the Klein-Nishina differential cross section $\mathrm{d}\sigma/\mathrm{d}\Omega$ will be used in the computations as discussed below. In any particular direction there is a unique monotonic relationship between $\omega'$ and $\omega$ and so a change of variables is possible yielding
\begin{equation} \label{maineq}
\frac{{\mathrm{d}^2 U_\gamma}}{{\mathrm{d}\omega ' \mathrm{d}\Omega }} = \frac{{\epsilon _0 c}}{{2\pi }}\left| {\widetilde E_x\left( {\omega \left( {\omega '} \right)} \right)} \right|^2 \frac{{\mathrm{d}\sigma }}{{\mathrm{d}\Omega }}\left[ {\frac{{\omega '}}{\omega }\frac{{\mathrm{d}\omega }}{{\mathrm{d}\omega '}}} \right].
\end{equation}

Next the fact that the electron bunch has non-zero emittance and energy spread must be accounted for. The easiest way to accomplish this task is, for every electron in the bunch: (i) Lorentz transform the incident wave packet to the electron rest frame, (ii)  Lorentz transform the propagation vector and polarization vector of the scattered wave into the electron frame, (iii) use the standard rest frame Klein-Nishina cross section to calculate the scattering from the electron in the lab frame, and (iv) sum the scattered energy of each individual electron. Therefore, one needs to evaluate and vary the scattering cross section slightly differently for each electron. The next task in this section is to give the general expression for the differential cross section for any possible kinematic condition for the electron.

For an electron at rest (beam rest frame), the Klein-Nishina differential scattering cross section for linearly polarized incident and scattered photons is
\begin{equation} \label{klnilincros}
\frac{{\mathrm{d}\sigma }}{{\mathrm{d}\Omega _b }} = \frac{{r_e^2 }}{4}\left( {\frac{{\omega '_b }}{{\omega _b }}} \right)^2 \left[ {\frac{{\omega '_b }}{{\omega _b }} + \frac{{\omega _b }}{{\omega '_b }} - 2 + 4\left( {\varepsilon _b  \cdot \varepsilon '_b } \right)^2 } \right],
\end{equation}
where $\omega_b$ and $\omega'_b$ are the incident and scattered radiation angular frequencies, respectively, with polarization 4-vectors $\varepsilon_b$ and $\varepsilon '_b$ . For future reference, the subscript {\it b} indicates a rest frame (beam frame) quantity, and throughout this paper polarization 4-vectors are of the form $\varepsilon=(0,\bm{\varepsilon})$. We use the metric with signature $(1,-1,-1,-1)$, so that the invariant scalar product of two 4-vectors $v_1^\mu$ and $v_2^\mu$ is $v_1\cdot v_2=v_1^0v_2^0-\bm{v}_1\cdot \bm{v}_2$.  The results of the scattering from each electron in a beam will be summed incoherently.

For notational convenience, the (implicit) summation over the individual electron coordinates is suppressed in the foregoing expressions. In our final summations to obtain observables in the lab frame, the relativistic factors $\bm{\beta}$ and $\gamma$ will apply to specific electrons. Straightforward Lorentz transformation from the rest frame of the individual beam electrons to the lab frame are made.  For example, the energy-momentum 4-vectors of the incident ($k^\mu=(\omega,\bm{k})$) and scattered ($k'^\mu=(\omega',\bm{k}')$) photons transform as
\begin{equation}
\begin{array}{l}
 \omega _b  = \gamma ( {1 - \bm{\beta}  \cdot \bm{\hat k}} )\omega  ,\\
 \bm{k}_b  =  - \gamma \bm{\beta} \omega  + \omega \bm{\hat k} + \omega \dfrac{( {\gamma  - 1})}{\beta ^2 }( {\bm{\beta}  \cdot \bm{\hat k}})\bm{\beta}  ,\\
 \omega '_b  = \gamma ( {1 - \bm{\beta}  \cdot \bm{\hat k}'})\omega ' ,\\
 \bm{k}'_b  =  - \gamma \bm{\beta} \omega ' + \omega '\bm{\hat k}' + \omega '\dfrac{( {\gamma  - 1})}{\beta ^2 }( {\bm{\beta}  \cdot \bm{\hat k}'})\bm{\beta},  \\
 \end{array}
\end{equation}
where $\bm{\hat{k}}=\bm{k}/|\bm{k}|$. Because the invariant scalar products $k\cdot k$ and $k'\cdot k'$ vanish, it readily follows that
\begin{equation}
\begin{array}{l}
 \bm{k}_b  \cdot \bm{k}_b  = \gamma ^2( {1 - \bm{\beta}  \cdot \bm{\hat k}})^2 \omega ^2  \to \bm{\hat k}_b  = \dfrac{ 1 }{\gamma(1 - \bm{\beta}  \cdot \bm{\hat k})} \left(
- \gamma \bm{\beta}  + \bm{\hat k} + \dfrac{(\gamma  - 1)}{\beta ^2 }(\bm{\beta}  \cdot \bm{\hat k})\bm{\beta}\right),\\
 \bm{k}'_b  \cdot \bm{k}'_b  = \gamma ^2(1 - \bm{\beta}  \cdot \bm{\hat k}')^2 \omega '^2  \to \bm{\hat k}'_b  = \dfrac{ 1 }{\gamma(1 - \bm{\beta}  \cdot \bm{\hat k}')} \left(
- \gamma \bm{\beta}  + \bm{\hat k}' + \dfrac{(\gamma  - 1)}{\beta ^2 }(\bm{\beta}  \cdot \bm{\hat k}')\bm{\beta}\right), \\
 \end{array}
\end{equation}
relating the unit propagation vectors in the electron rest frame to those in the lab frame.

The potential 4-vectors for the incident and scattered photons in the lab frame are $(0,\bm{\varepsilon})Ae^{i(\omega t-\bm{k}\cdot\bm{x})}$ and $(0,\bm{\varepsilon}')A'e^{i(\omega' t-\bm{k}'\cdot\bm{x})}$. Using the 4-vector transformation formula and performing a gauge transformation to eliminate their zeroth components lead to $(0,\bm{\varepsilon}_b)Ae^{i(\omega_b t_b-\bm{k}_b\cdot\bm{x}_b)}$ and $(0,\bm{\varepsilon}'_b)A'e^{i(\omega'_b t_b-\bm{k}'_b\cdot\bm{x}_b)}$ where
\begin{equation}
\begin{array}{l}
 \bm{\varepsilon} _b  = \gamma \left( {\bm{\beta}  \cdot \bm{\varepsilon} } \right)\bm{\hat k}_b  + \bm{\varepsilon}  + \dfrac{(\gamma  - 1)}{\beta ^2 }\left( {\bm{\beta}  \cdot \bm{\varepsilon} } \right)\bm{\beta} , \\
 \bm{\varepsilon} '_b  = \gamma \left( {\bm{\beta}  \cdot \bm{\varepsilon} '} \right)\bm{\hat k}'_b  + \bm{\varepsilon} ' + \dfrac{(\gamma  - 1)}{\beta ^2 }\left( {\bm{\beta}  \cdot \bm{\varepsilon} '} \right)\bm{\beta} . \\
 \end{array}
\end{equation}
Because the beam frame polarization vector is linearly related to the lab frame polarization vector, equivalent expressions apply to the transformation of the complex polarization vectors needed for describing circular or elliptical polarization which will be used in Section \ref{sec:circular}. To evaluate the Klein-Nishina cross section, one can use the relation
\begin{equation}
\bm{\varepsilon} _b  \cdot \bm{\varepsilon} '_b
  = \bm{\varepsilon}  \cdot \bm{\varepsilon} ' + \frac{(\bm{\beta}  \cdot \bm{\varepsilon})( \bm{\hat k} \cdot \bm{\varepsilon} ')}{(1 - \bm{\beta}  \cdot \bm{\hat k})}
+ \frac{(\bm{\beta}  \cdot \bm{\varepsilon} ')(\bm{\hat k}' \cdot \bm{\varepsilon} )}{( {1 - \bm{\beta}  \cdot \bm{\hat k}'})} + \gamma ^2(\bm{\beta}  \cdot \bm{\varepsilon})(\bm{\beta}  \cdot \bm{\varepsilon} ')(\bm{\hat k}_b  \cdot \bm{\hat k}'_b  - 1).
\end{equation}
Rewriting in terms of the 4-scalar product yields
\begin{equation} \label{pdef}
\varepsilon _b  \cdot \varepsilon '_b  = \varepsilon  \cdot \varepsilon ' - \frac{( p_i  \cdot \varepsilon )(k \cdot \varepsilon ')}{p_i  \cdot k} - \frac{(p_i  \cdot \varepsilon ')(k' \cdot \varepsilon )}{p_i  \cdot k'} + \frac{(p_i  \cdot \varepsilon )(p_i  \cdot \varepsilon ')(k \cdot k')}{(p_i  \cdot k)(p_i  \cdot k')} \equiv P\left( {\varepsilon ,\varepsilon '} \right),
\end{equation}
where $p_i$ is the 4-momentum of the incident electron. Note that because the 4-vectors $p_i$, $k$, and $k'$ are real, one has $P(\varepsilon_1,\varepsilon^*_2)=
P^*(\varepsilon^*_1,\varepsilon_2)$ and $P(\varepsilon^*_1,\varepsilon^*_2)=
P^*(\varepsilon_1,\varepsilon_2)$.

The standard calculation of the lab frame phase-space factor yields the generalized Compton formula
\begin{equation} \label{compton}
\omega ' = \frac{\omega (1 - \bm{\beta}  \cdot \bm{\hat k})}
{1 - \bm{\beta}  \cdot \bm{\hat k}' + (\hbar\omega /\gamma mc^2)( 1 - \bm{\hat k} \cdot \bm{\hat k}')}
\end{equation}
and the expression for the lab frame cross section is
\begin{equation} \label{crosssect1}
\frac{{\mathrm{d}\sigma }}{{\mathrm{d}\Omega }}
= \frac{r_e^2 }{4\gamma ^2 (1 - \bm{\beta}  \cdot \bm{\hat k} )^2 }
\left( \frac{\omega '}{\omega } \right)^2
\left[
\frac{\omega '(1 - \bm{\beta}  \cdot \bm{\hat k}')}
{\omega (1 - \bm{\beta}  \cdot \bm{\hat k} )}
+ \frac{\omega (1 - \bm{\beta}  \cdot \bm{\hat k} )}
{\omega '(1 - \bm{\beta}  \cdot \bm{\hat k}' )}
- 2 + 4 [P(\varepsilon ,\varepsilon ')]^2
\right].
\end{equation}
When $\bm{\beta}=0$ this expression obviously reduces to the rest frame result, and applies when a linearly polarized laser beam is scattered by an unpolarized electron beam. It captures the dependence on linear polarization in both the initial and final states. The expression in Eq.~(\ref{crosssect1}) is a modification of a result found in Ref.~\cite{ovv}. Because the cross section is written here in terms of the incident electron and photon momenta, and in most Compton sources the recoil electron is not detected, this form is most convenient for integrating over the beam electron and incident laser photon distributions.

 In our numerical calculations it is assumed that the polarization of the scattered photons is not observed. In this case the total cross section is the sum of the cross sections for scattering into the two orthonormal final state polarization vectors. The polarization sums may be replaced by scalar products as usual \cite{pessch}. Defining
\begin{equation} \label{pvectord}
P^\mu( \varepsilon) \equiv \varepsilon ^\mu
  - \frac{p_i  \cdot \varepsilon }{p_i  \cdot k}k^\mu
  - \frac{k' \cdot \varepsilon }{p_i  \cdot k'}p_i^\mu
 + \frac{(p_i  \cdot \varepsilon)(k \cdot k')}{(p_i  \cdot k)(p_i  \cdot k')}p_i^\mu,
\end{equation}
one can write the scalar product in Eq.~(\ref{pdef}) as $P(\varepsilon,\varepsilon')=P^\mu(\varepsilon)\varepsilon'_\mu$. Because $P^\mu(\varepsilon) k'_\mu=0$, one has
\begin{equation} \label{polsum}
\begin{array}{rl}
- P^\mu( \varepsilon) P_\mu(\varepsilon)&=P(\varepsilon ,\varepsilon '_1)P(\varepsilon  ,\varepsilon '_1) + P( \varepsilon ,\varepsilon '_2)P(\varepsilon  ,\varepsilon '_2)   \\
  &= 1 - m^2 c^2 \left[\dfrac{(k' \cdot \varepsilon)(k' \cdot \varepsilon)}{(p_i  \cdot k')^2 }
 - 2\dfrac{(p_i  \cdot \varepsilon)(k' \cdot \varepsilon)}{(p_i  \cdot k)(p_i  \cdot k')^2 }(k \cdot k')
+ \dfrac{(p_i  \cdot \varepsilon)(p_i \cdot\varepsilon )}{(p_i  \cdot k)^2(p_i  \cdot k')^2 }(k \cdot k')^2 \right]  ,\\
 \end{array}
\end{equation}
for any two orthonormal polarization vectors $\varepsilon'_1$  and $\varepsilon'_2$ orthogonal to the propagation vector  $k'$. The differential cross section summed over the final polarization is
\begin{equation} \label{cross2}
\frac{\mathrm{d}\sigma }{\mathrm{d}\Omega } = \frac{r_e^2 }{2\gamma ^2(1 - \bm{\beta}  \cdot \bm{\hat k})^2 }
\left(\frac{\omega '}{\omega }\right)^2  \left[
 \frac{\omega '(1 - \bm{\beta}  \cdot \bm{\hat k}')}
{\omega (1 - \bm{\beta}  \cdot \bm{\hat k})}
+ \frac{\omega(1 - \bm{\beta}  \cdot \bm{\hat k})}
{\omega '(1 - \bm{\beta}  \cdot \bm{\hat k}')}
 - \frac{2m^2 c^2 }{(p_i  \cdot k')^2 }\left(k' \cdot \varepsilon - \frac{(p_i  \cdot \varepsilon)}{(p_i  \cdot k)}k \cdot k' \right)^2 \right].
\end{equation}
This differential cross section, inserted in Eq.~(\ref{maineq}), is used to calculate the spectrum of the scattered radiation for a single electron. The total scattered energy is obtained by summing the spectra, each generated using the relativistic factors for each electron. A more general expression, correctly accounting for the circularly or elliptically polarized photons is presented in Section \ref{sec:circular}. It should be noted that the individual and summed differential cross sections in Eqs.~(\ref{crosssect1}) and (\ref{cross2}) are somewhat different from those reported in Ref.~\cite{ELI1}.

In order to determine the overall scale of the spectrum expected in the numerical results, and to provide contact with previous calculations, it is worthwhile to take the Thomson limit of these expressions. At low incident frequency, the recoil term involving the electron mass in Eq.~(\ref{compton}) becomes negligible. The relationship between incident and scattered frequency is then
\begin{equation} \label{thomson}
\omega ' = \frac{\omega(1 - \bm{\beta}  \cdot \bm{\hat k})}{1 - \bm{\beta}  \cdot \bm{\hat k}' },
\end{equation}
and the expression for the lab frame cross section is
\begin{equation} \label{crosssect4}
\frac{\mathrm{d}\sigma }{\mathrm{d}\Omega } = \frac{r_e^2 }{\gamma ^2(1 - \bm{\beta}  \cdot \bm{\hat k}')^2} \left[1 - \frac{m^2 c^2 }{(p_i  \cdot k')^2 } \left(k' \cdot \varepsilon - \frac{(p_i  \cdot \varepsilon )}{(p_i  \cdot k)}k \cdot k' \right)^2 \right].
\end{equation}

For an electron moving on the $z$-axis and backscattering with an $x$-polarized incident photon moving anti-parallel, the differential cross section simplifies to
\begin{equation}
\frac{{\mathrm{d}\sigma }}{{\mathrm{d}\Omega }} = \frac{{r_e^2 }}{{\gamma ^2 \left( {1 - \beta\cos\theta} \right)^2 }}\left[\frac{(1-\beta\cos\theta)^2\sin^2\phi+(\cos\theta-\beta)^2\cos^2\phi
}{(1-\beta\cos\theta)^2} \right],
\end{equation}
consistent with Eq.~(\ref{thomlt}) above.

Generally speaking, from an experimental point of view, it is most interesting to know the number of scattered photons per unit scattered energy. To determine this quantity note that, by the convolution theorem, the Fourier transform of the normalized vector potential function $a_m(t)\cos(\omega_0t)$ is
\begin{equation}
\tilde a(\omega)=\frac{{\tilde a_m(\omega-\omega_0)+\tilde a_m(\omega+\omega_0)}}{{2}}.
\end{equation}
Therefore, after completing the trivial integrations over $\phi$,
\begin{equation} \label{maineqggg}
\frac{{\mathrm{d}U_\gamma}}{{\mathrm{d}\omega'}} = \frac{{r_e^2\epsilon _0 c\pi}}{{8\pi }}\left(\frac{{mc}}{{e}} \right)^2 \int \left(\frac {{\omega'(1-\beta\cos\theta)}}{{(1+\beta)}}\right)^2 \left| \tilde a_m\left( \frac {{\omega'(1-\beta\cos\theta)}}{{(1+\beta)}} \right) \right|^2 \left[\frac{(1-\beta\cos\theta)^2+
(\cos\theta-\beta)^2}{{\gamma^2(1-\beta\cos\theta)^4}} \right]\mathrm{d}(\cos\theta).
\end{equation}

For an amplitude function slowly varying on the time scale of the oscillation, the Fourier transform of $a_m$ is highly peaked as a function of $\cos\theta$. Changing variables, using Parseval's theorem to evaluate the frequency integral, and collecting constants yields \cite{kjkim}
\begin{equation} \label{maineqggggg}
\frac{{\mathrm{d}U_\gamma}}{{\mathrm{d}E_\gamma}} \doteq \frac{{(1+\beta)c\,\alpha\,\pi}}{{4\beta\lambda}} \frac{{E_\gamma}}{{E_{\gamma,\mathrm{max}}}}  \left[\frac{(1-\beta\cos\theta)^2+
(\cos\theta-\beta)^2}{(1-\beta\cos\theta)^2} \right]
\int_{-\infty}^\infty a_m^2(t)\, \mathrm{d}t ,
\end{equation}
where $E_\gamma=\hbar\omega'$, $E_{\gamma,{\mathrm{max}}}=(1+\beta)^2\gamma^2\hbar\omega_0$ is the Compton edge maximum energy emitted in the forward direction, $\alpha$ is the fine structure constant, and $\lambda=2\pi c/\omega_0$ is the incident laser wavelength. The (equal) contributions from both positive and negative frequencies in the Fourier transform of the field are accounted in Eq.~(\ref{maineqggggg}).

As a final step, replacing $\cos\theta$ by $E_\gamma$, one obtains
\begin{equation} \label{cedge}
\frac{{\mathrm{d}U_\gamma}}{{\mathrm{d}E_\gamma}} \doteq \frac{{(1+\beta)c\,\alpha\,\pi}}{{4\beta^3\lambda}}\frac{{E_\gamma}}{{E_{\gamma,\mathrm{max}}}} \left[\beta^2 +\left(\frac{{(1+\beta)E_\gamma}}{{E_{\gamma,\mathrm{max}}}}-1\right)^2 \right] \int_{-\infty}^\infty a_m^2(t)\, \mathrm{d}t .
\end{equation}
The number density of all photons produced as a function of scatterred energy is easily found from this equation simply by dividing by $E_\gamma$. The number distribution is precisely parabolic in the Thomson limit, with minimum
value of $\beta^2$ at $E_\gamma=(1+\beta)\gamma^2\hbar\omega_{laser}$, also the
average energy of all photons. The number density
grows to a value $2\beta^2$ at both the
Compton edge in the forward direction, and in the backward direction where the laser frequency is not Doppler shifted.

Equation (\ref{cedge}) provides an excellent check of the scale for the results from the numerical technique. When the electron emittance and energy spread vanish, and one takes the long pulse limit, the height of the energy spectrum is
\begin{equation}
\frac{{\mathrm{d}U_\gamma}}{{\mathrm{d}E_\gamma}} \doteq \frac{{ c\,\alpha\,\pi}}{{\lambda}} \int_{-\infty}^\infty a_m^2(t)\, \mathrm{d}t ,
\end{equation}
and the height of the number spectrum is
\begin{equation}
\frac{{\mathrm{d}N_\gamma}}{{\mathrm{d}E_\gamma}} \doteq \frac{{c\,\alpha\,\pi}}{{\lambda E_{\gamma,\mathrm{max}}}} \int_{-\infty}^\infty a_m^2(t)\, \mathrm{d}t ,
\end{equation}
at the Compton edge.

Throughout this work the plane wave approximation is used. However, at the expense of a greater number of computations for each electron, it is possible to capture three dimensional effects in the photon pulses using our general approach. The main adjustments are to modulate the vector potential because of the electron orbit through the three dimensional photon pulse structure and to include the arrival time variation of the individual electrons. A common incident photon spectrum for all of the electrons is no longer possible  \cite{Harvey}.  Our present intent is to undertake a more general code including such improvements and to publish calculations, including benchmarks, in a future publication. Presently, we anticipate that there may be computation time advantages from pursuing spectrum calculations using this approach compared to straight simulation calculations such as CAIN \cite{CAIN}.

\section{Numerical Method} \label{sec:nummeth}

In the previous section, we derived the general expression for energy density per solid angle for the Compton scattered photons from
a laser beam by one electron, Eq.~(\ref{maineq}).
The first non-constant term in Eq.~(\ref{maineq}) quantifies the electric field produced
by the laser. The remaining terms are general and independent on the specifics of the
experimental setup---the second non-constant term gives the probability that a photon is
scattered into a given solid angle $\mathrm{d}\Omega$ and the third is the relativistic relationship
between the frequencies of the incident and scattered radiation. Each of the non-constant
terms depends on the scattered angular frequency $\omega'$ and the solid angle
$\mathrm{d}\Omega = \mathrm{d}\phi \mathrm{d}\cos\theta$.

In order to compute the energy spectrum captured by a detector in a laboratory, the energy density per solid angle should be integrated over the solid angle of the aperture for a representative
sample of particles from the electron beam. The resulting energy density spectrum
for each electron is
\begin{equation} \label{eqUE}
\frac{{\mathrm{d}U_1}}{{\mathrm{d}\omega'}} =
\frac{\epsilon_0c}{2\pi}
\int_{0}^{2\pi} \mathrm{d}\phi
\int_{\cos \theta_a}^{1}
\left| {\widetilde E}\left(\omega(\omega')\right)\right|^2
\frac{\mathrm{d} \sigma}{\mathrm{d} \Omega}
\left[\frac{\omega'}{\omega} \frac{\mathrm{d}\omega}{\mathrm{d}\omega'}\right]\mathrm{d}(\cos\theta),
\end{equation}
where $\theta_{a}$ is the semi-angle of the aperture and the subscript ``$1$" denotes that the quantity is due to scattering off a
single electron. Although essentially the same quantity is computed numerically by a somewhat different procedure in Ref.~\cite{ELI1}, we have observed that integrating with $\cos\theta$ as the independent variable markedly increases the precision of the numerical results. The equivalent number density of the spectrum
is given by
\begin{equation}
\frac{\mathrm{d}N_1}{\mathrm{d}\omega'} =
\frac{1}{\hbar \omega'}
\frac{\mathrm{d}U_1}{\mathrm{d}\omega'}.
\end{equation}
Only in the limiting case when the laser width approaches infinity and the pulse tends
to a continuous wave (CW) is the integration over $\cos\theta$ analytically tractable. In every other case, numerical
integration of Eq.~(\ref{eqUE}) is required.

For a representative subset of $N_{p}$ particles from an electron beam distribution
\begin{equation}
f({\bm{p}}) = \sum_{i=1}^{N_{p}} \delta ({\bm{p}} - {\bm{p}}_i),
\end{equation}
where ${\bm{p}} = (p_x, p_y, p_z)$, the total energy density and number density spectra
per electron are, respectively,
\begin{equation} \label{eqNE}
\frac{\mathrm{d}U}{\mathrm{d}\omega'} = \frac{1}{N_{p}}
\sum_{i=1}^{N_{\rm p}} \frac{\mathrm{d}U_1}{\mathrm{d}\omega'}({\bm{p}}_i), \hskip15pt
\frac{\mathrm{d}N}{\mathrm{d}\omega'} = \frac{1}{N_{p}}
\sum_{i=1}^{N_{\rm p}} \frac{\mathrm{d}N_1}{\mathrm{d}\omega'}({\bm{p}}_i).
\end{equation}
It is instructive to recall that the accuracy of results produced from a random sample on
$N_p$ particles is proportional to $1/\sqrt{N_p}$. Therefore, for example, for a 1\% accuracy
in computed spectra, an average over 10,000 points representing the underlying electron
beam distribution is needed.

We implement the numerical integration of Eq.~(\ref{eqNE}) in the {\tt Python} scripting language
\cite{python}. The two-dimensional integration is performed using the {\tt dblquad} routine
from the {\tt scipy} \cite{scipy} scientific {\tt Python} library. {\tt dblquad} performs a
two-dimensional integration by computing two nested one-dimensional
quadratures using an adaptive, general-purpose integrator based on the {\tt qag}
routine from {\tt QUADPACK} \cite{quadpack}. This general-purpose integrator performs
well even for moderate-to-highly peaked electric fields $\sigma\le 200$, where $\sigma$ is the ratio of the length of the field falloff and the wavelength. For $\lambda=800$ nm this condition
requires the laser pulse duration to be shorter than $\tau = \sigma\lambda/c \approx 0.5$ ps.
However, as the laser pulse duration increases beyond this approximate range, the
electric field becomes extremely peaked, and the general purpose integrator
{\tt qag} can no longer handle this computation without occasionally generating spurious
results. It simply is not designed to handle such extreme integrand behavior.
Efforts are currently underway to replace the {\tt qag} integrator with a state-of-the-art,
intrinsically multidimensional, adaptive algorithm optimized to run on CPU and
GPU platforms \cite{a13a,a13b}.

Although the summation over electrons will be performed with an actual computer-generated distribution from the ODU Compton source design, it is worthwhile to summarize some facts about the numerical distributions for the electrons used in test cases to check the calculation method. The electron momenta are generated as
\begin{eqnarray}
p_x & = & p \sqrt{\frac{\varepsilon_x}{\beta_x^{*}}} \; \delta(0,1), \nonumber \\
p_y & = & p \sqrt{\frac{\varepsilon_y}{\beta_y^{*}}} \; \delta(0,1), \\
p_z & = & \sqrt{[p + \sigma_p \; \delta(0,1)]^2 - p_x^2 - p_y^2}, \nonumber
\end{eqnarray}
where $p = \sqrt{p_x^2+p_y^2+p_z^2}$ is the magnitude of the total momentum,
 $\delta(0,1)$ is a Gaussian-distributed random variable with zero mean
and unit variance and $\sigma_p$ is the standard deviation in the total momentum:
\begin{equation}
\sigma_p = \sqrt{\left[E_0\left(1+\frac{\sigma_E}{E}\right)\right]^2 - m^2 c^4}.
\end{equation}
Neglecting the small difference between the magnitude of the momentum and the $z$-component of the momentum, $\sigma_{\theta_x}=\varepsilon_x/\beta_x^*$ and $\sigma_{\theta_y}=\varepsilon_y/\beta_y^*$ are therefore the {\it rms} spread in beam transverse angles and $\sigma_p$ is the relative longitudinal momentum spread. Using the relativistic energy-momentum relation in the ultrarelativistic limit and the usual statistical averaging, one obtains
\begin{equation}
\frac {{\sigma_{E_{e}}}}{{E_{e}}}=\sqrt{\sigma_p^2(1-\sigma_{\theta_x}^2-\sigma_{\theta_y}^2)+\sigma_{\theta_x}^4/2 + \sigma_{\theta_y}^4/2}
\end{equation}
for the relative energy spread of the electrons generated including all terms up to fourth order in the small quantities $\sigma_i$. Notice that as the beam emittance changes there is a change in the energy spread generated at the same time.

The sheer amount of computation required to obtain the spectra with appropriate
experimental (number of scattered energies $E_{\gamma}$) and statistical
(number of electrons sampling the distribution) resolution is substantial. This problem
was alleviated by parallelizing the computation to efficiently run on multicore platforms.
For this purpose, {\tt Python}'s {\tt multiprocessing} library was used. The code is available
upon request.

The code takes as input parameters of the inverse Compton scattering: (i) the
properties of the electron beam (energy $E_e$, energy spread $\sigma_{E}$,
horizontal emittance $\varepsilon_x$, vertical emittance $\varepsilon_y$, total charge $Q$);
(ii) the properties of the laser beam (energy $E_l$, energy spread $\sigma$,
amplitude of the normalized vector potential $a_0$); (iii) the shape of the
laser beam (i.e., Gaussian, hard-edge, etc.); (iv) the properties of the aperture
(size and location) and (v) the resolution of the simulation
(the number of scattered radiation energies at which the spectrum is computed
and the number of particles sampling the electron beam particle distribution).
The output is the numerically computed scattered radiation spectrum.

\section{Model Validation and Benchmarking} \label{sec:bmv}

The generalized Compton formula for the angular frequency $\omega'$ in Eq.~(\ref{compton}) can be written in more explicit form in the lab frame as
\begin{equation}
\omega'(\omega)=\frac{\omega(1+\beta p_{z}/|\bm{p}|)}
{1-(\beta/|\bm{p}|)(p_{x}\sin\theta\cos\phi+p_{y}\sin\theta\sin\phi+p_{z}\cos\theta)+(\hbar\omega/\gamma mc^{2})\left(1+\cos\theta\right)},
\end{equation}
where $\beta=|\bm{\beta}|$. The greatest gain in angular frequency scattered is from an electron beam incident along $\hat{z}$ and at $\theta=0$, i.e., near the $z$-axis. We calculate the expected spectrum incident upon a circular, on-axis sensor aperture of radius $R$ using Eq.~(\ref{eqUE}). This geometry will be assumed in all the cases we consider.

\begin{table}
\caption[Scattering Parameters for Aperture Plots]{Electron Beam and Incident Laser Pulse Parameters Used in Through-Aperture Spectra Calculation.}
\label{Tab: Aperpar}
\vspace{12pt.}
\begin{center}
\begin{tabular}{lcc}\hline\hline
Parameter & Symbol & Value \\
\hline
Electron beam energy& $E_{b}$ & 500 MeV  \\
Peak normalized vector potential& $a_{0}$ & 0.026\\
Incident photon spread parameter&$ \sigma$& 50 \\
Peak laser pulse wavelength & $\lambda$ & 800 nm\\
Aperture distance from collision & $L$ & 60 m \\
Horizontal emittance & $\varepsilon_{x}$ & 0.05 nm\thinspace rad\\
Vertical emittance & $\varepsilon_{y}$ & 0 nm\thinspace rad\\
Electron energy spread & $\sigma_{E_{e}}/E_{e}$ & $2\times10^{-3}$ \\ \hline\hline
\end{tabular}
\end{center}\label{Tab: par}
\end{table}

\begin{figure}
\begin{center}
\includegraphics[width=3.22in]{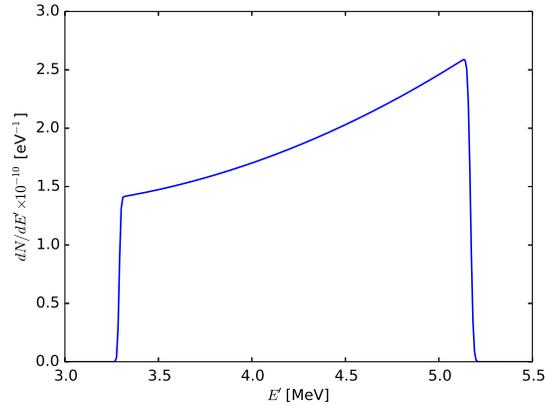}
\caption{\small{
The number density of the energy spectrum of a Compton gamma-ray beam produced
by the head-on collision of a 466 MeV electron with a 789 nm laser beam,
as in Fig.~2 in Ref.~\cite{slrtw09}. A collimation aperture with radius $R$ of 50 mm is
placed a distance $L$ of 60 m downstream from the collision point ($\theta_a=\tan^{-1}(R/L)$). Here a Gaussian laser pulse
with $\sigma=50$ is used, while the laser in
Ref.~\cite{slrtw09} is a CW.
}}
\label{fig:fig2}
\end{center}
\end{figure}


It is evident that for a CW laser beam
the Fourier transform of the electric field is simply a delta function. A pulsed laser
model, in contrast, leads to a distribution in frequencies, with an intrinsic
energy spread. The CW model, while
useful in making the resulting spectra analytically tractable, does not allow for
studying the effects of the pulsed nature of the laser beam. For instance,
the relative importance of the energy spreads of the two colliding beams on the
shape of the spectra of the backscattered radiation can only be addressed with
a pulsed model. Here we consider a general pulsed structure of the incident laser beam.

The electric field can either be computed from the initially
prescribed shape of the laser pulse or specified directly.
In this paper and in the code, we provide one example of each:
(i) electric field computed from the Gaussian laser pulse and
(ii) electric field directly specified to be a hard-edge pulse, modeling
a flat laser pulse.

Fourier transforming the Gaussian laser pulse
\begin{equation}
A_x(t) = A_0 \exp\left(-\frac{c^2t^2}{2(\sigma\lambda)^2} \right)
\cos\left(\frac{2\pi ct}{\lambda}\right),
\end{equation}
yields the transformed electric field
\begin{equation}
{\widetilde E_x} (\omega) = -i\frac{\omega A_{0}\sigma\lambda}{c}\sqrt{\frac{\pi}{2}}
\left[
\exp\left({-\frac{\left(\sigma\lambda\right)^{2}}{2c^2}\left(\omega-\frac{2\pi c}
{\lambda}\right)^{2}}\right)
+\exp\left({-\frac{\left(\sigma\lambda\right)^{2}}{2c^2}\left(\omega+\frac{2\pi c}
{\lambda}\right)^{2}}\right)
\right],
\end{equation}
where $A_0$ is the maximum amplitude of the vector potential, and we denote the normalized vector potential by $a_0 = eA_0/m c$. In the limit of $\sigma \to \infty$, the laser transitions from the pulsed to
CW nature, the electric field becomes two $\delta$-functions at $\omega=\pm 2\pi c/\lambda$
and earlier results such as those in
Fig.~2 of Ref.~\cite{slrtw09} are recovered. Figure~\ref{fig:fig2} shows the
number density of the energy spectrum for a pulsed very-wide
wave with $\sigma=50$. The perfect agreement of the overall scale in the two plots, computed two different ways, validates our numerical approach to this problem. In addition, the calculation captures the main effect expected from frequency spread in the incident laser: both sharp edges in the spectrum should be washed out so that the transition happens on a relative frequency scale equal to the relative frequency spread in the pulse.

A hard-edge laser pulse, modeling a flat laser pulse, is given by
\begin{equation}
A_x(t) = A_0
\cos(2\pi ct/\lambda)
\left[
\Theta(t+N\lambda/2c) -
\Theta(t-N\lambda/2c)
\right],
\end{equation}
where $\Theta(x)$ is the Heaviside step function, and $N$ is the
number of periods of the laser within the hard-edge pulse.
The corresponding transformed electric field is
\begin{equation}
{\widetilde E_x} (\omega) = -i\omega A_{0}
\left[
\frac{\sin((\omega -2\pi c/\lambda)N\lambda/2c)}{\omega -2\pi c/\lambda}
+ \frac{\sin((\omega +2\pi c/\lambda)N\lambda/2c)}{\omega +2\pi c/\lambda}
\right].
\end{equation}
Again, in the limit of $N \to \infty$, the laser transitions from the pulsed to
CW nature, the electric field reduces to two delta functions
and the earlier results are recovered. An identical plot to our Fig.~\ref{fig:fig2}
and Fig.~2 of Ref.~\cite{slrtw09} is produced for $N=50$.

\begin{figure}
\begin{center}
\includegraphics[width=3.22in]{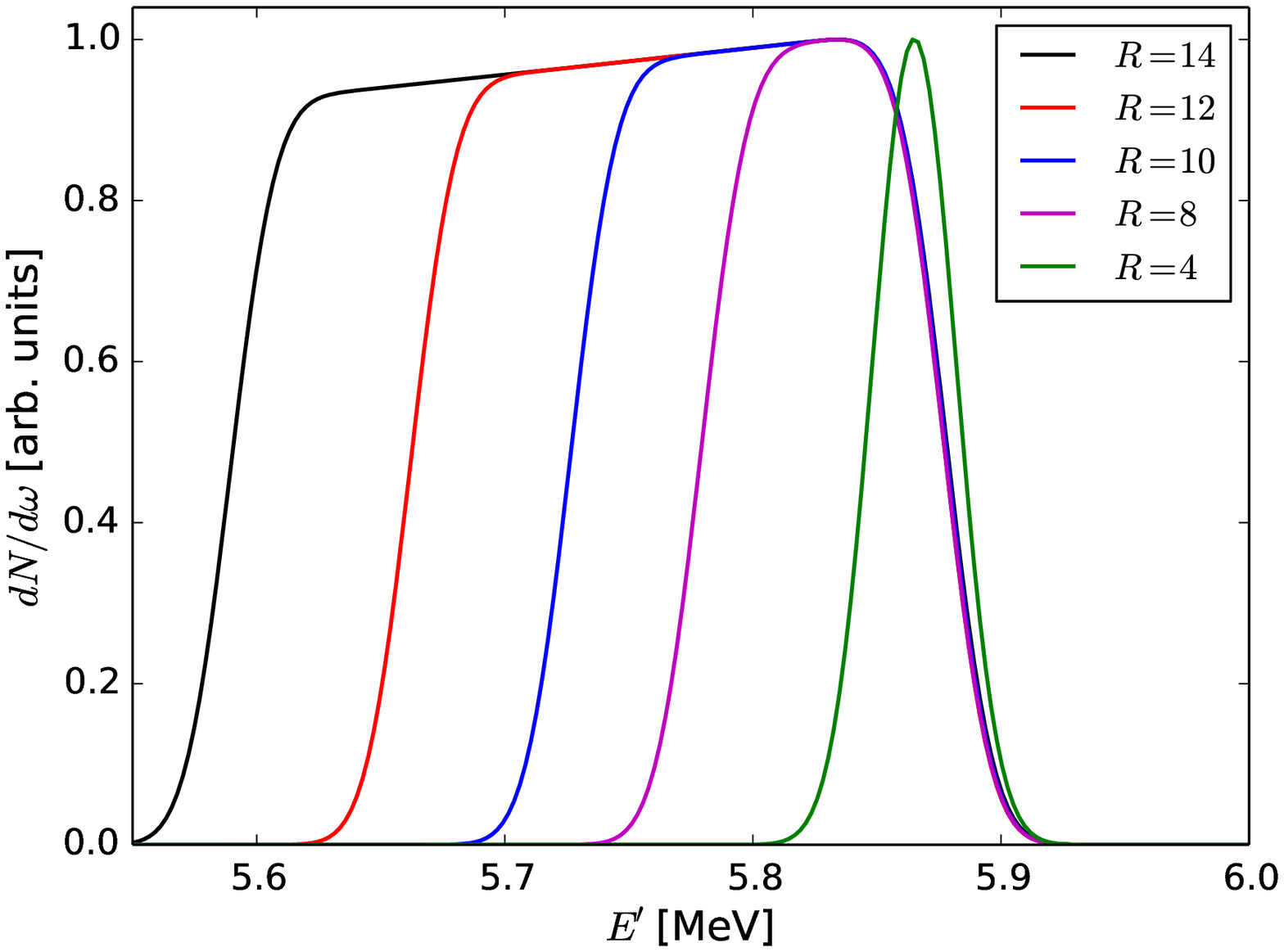}
\includegraphics[width=3.22in]{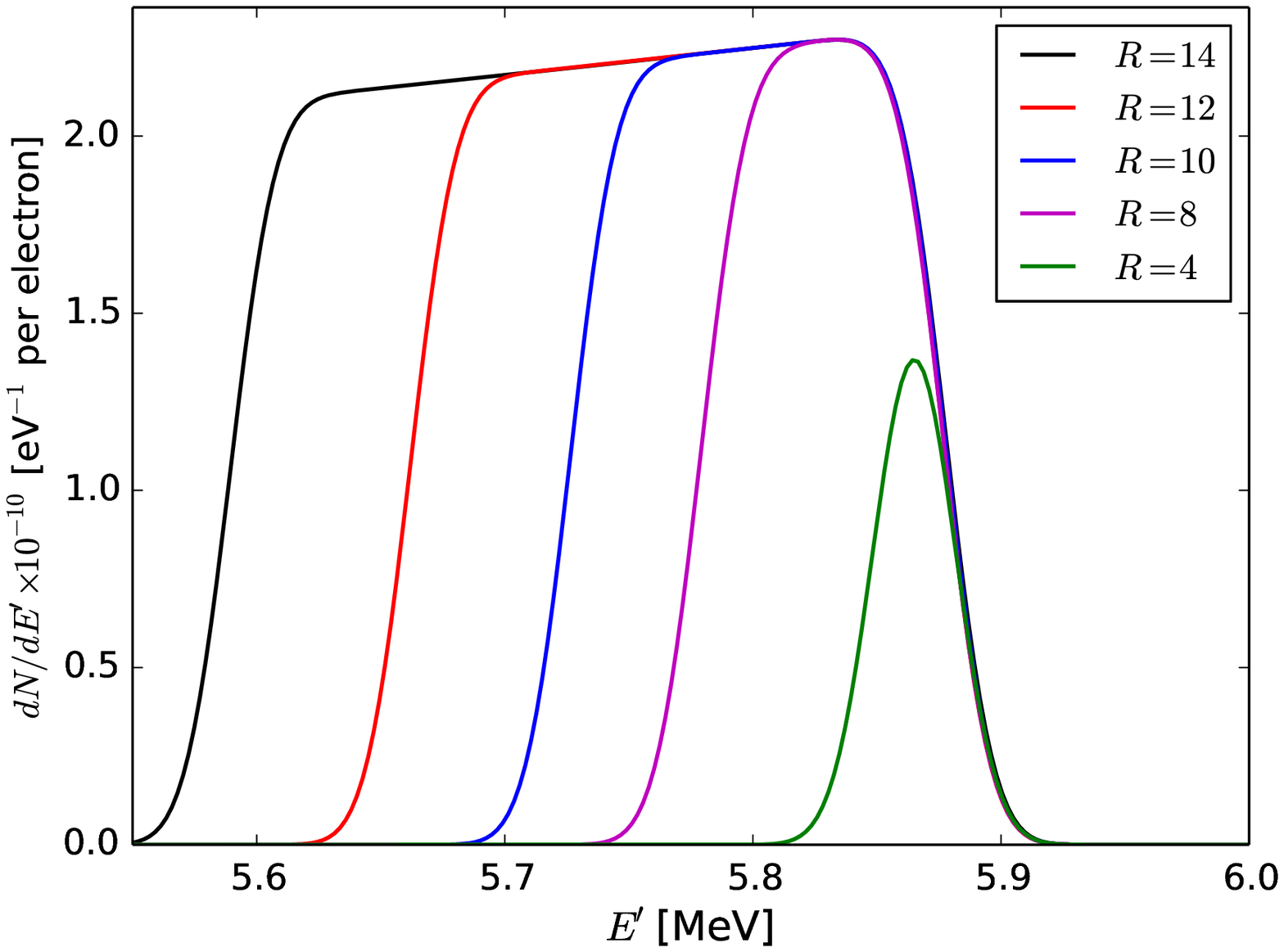}
\caption{\small{
The number density of the energy spectrum of a Compton gamma-ray beam produced
by the head-on collision of a 500 MeV electron with a 800 nm pulsed laser beam, for
different radii $R$ of the collimation point (in mm), as in Fig.~3 in Ref.~\cite{slrtw09}.
The aperture is 60 m downstream from the collision point.
The horizontal emittance and energy spread of the electron beam are
held constant at 0.05 nm rad and $2 \times 10^{-3}$, respectively.
Here a Gaussian laser pulse
with $\sigma=50$ is used, while the laser in
Ref.~\cite{slrtw09} is a CW.
Each curve is generated by averaging 400 electrons sampling the prescribed
distribution.
Left panel: Spectra scaled to their respective peak values (compare
with Fig.~3(b) of Ref.~\cite{slrtw09}). Right panel: Spectra in physical units.}}
\label{fig:fig3}
\end{center}
\end{figure}

\begin{figure}
\begin{center}
\includegraphics[width=3.22in]{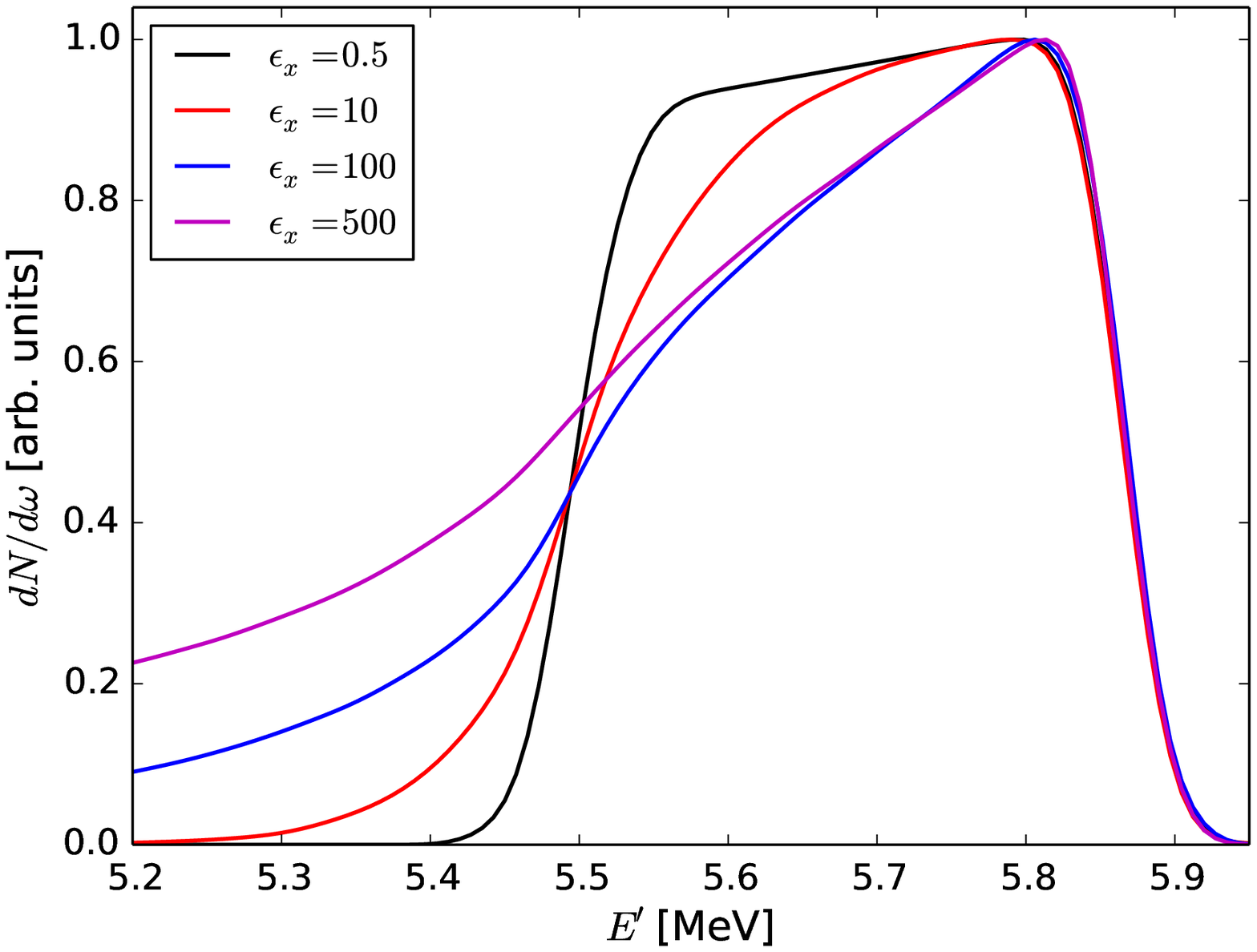}
\includegraphics[width=3.22in]{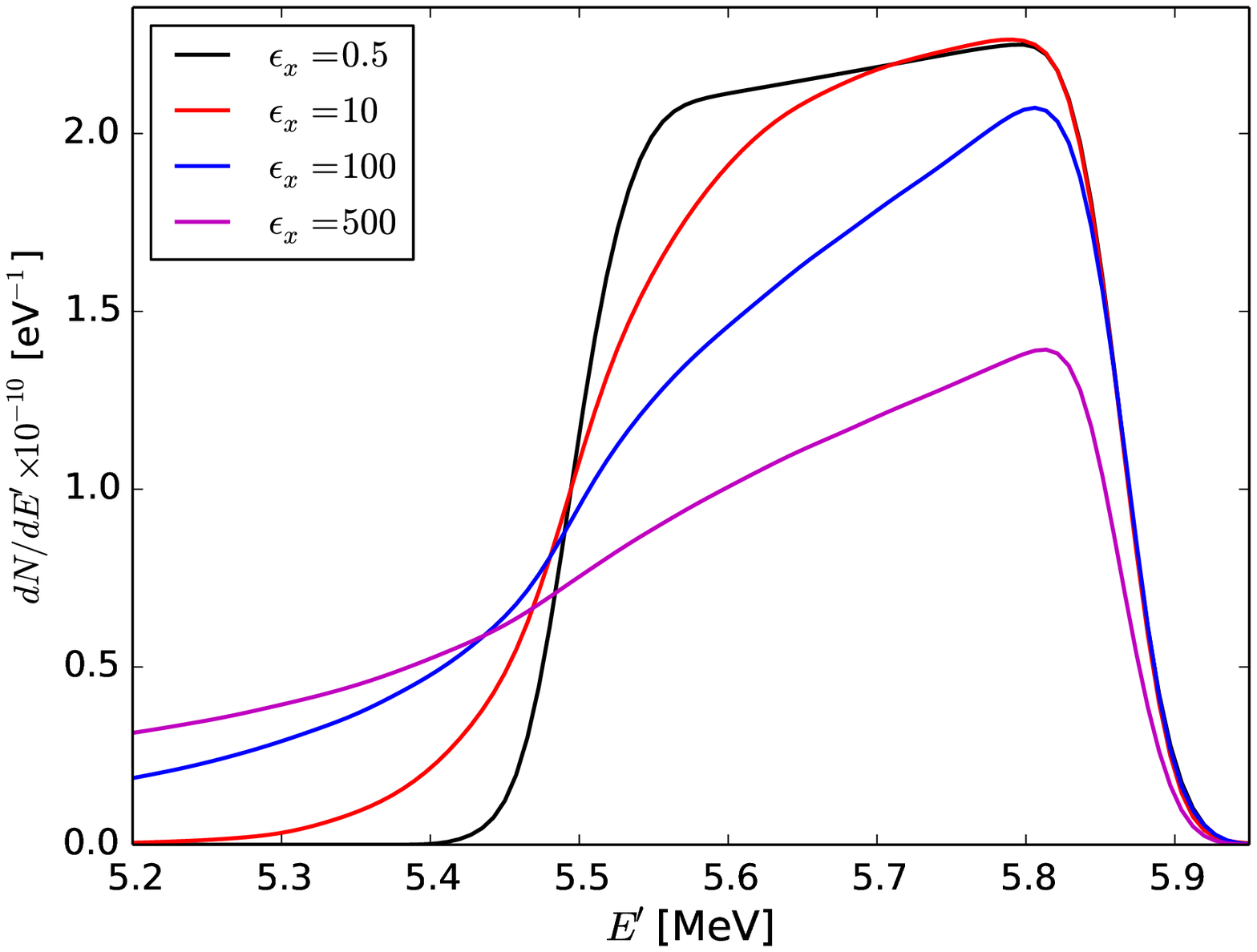}
\caption{\small{
The number density of the energy spectrum of a Compton gamma-ray beam produced
by the head-on collision of a 500 MeV electron with a 800 nm pulsed laser beam, for
different horizontal emittances $\epsilon_x$ as in Fig.~4(a) in Ref.~\cite{slrtw09}.
The laser is collimated by an aperture with radius of 16 mm, placed 60 m downstream
from the collision point. The relative energy spread of the electron beam $\sigma_E$
is held constant at $2 \times 10^{-3}$.
Here a Gaussian laser pulse with $\sigma=50$ is used,
while the laser in Ref.~\cite{slrtw09} is a CW.
Each curve is generated by averaging 400 electrons sampling the prescribed distribution.
Left panel: Spectra scaled to their respective peak values (compare
with Fig.~4(a) of Ref.~\cite{slrtw09}). Right panel: Spectra in physical units.}}
\label{fig:fig4}
\end{center}
\end{figure}

\begin{figure}
\begin{center}
\includegraphics[width=3.22in]{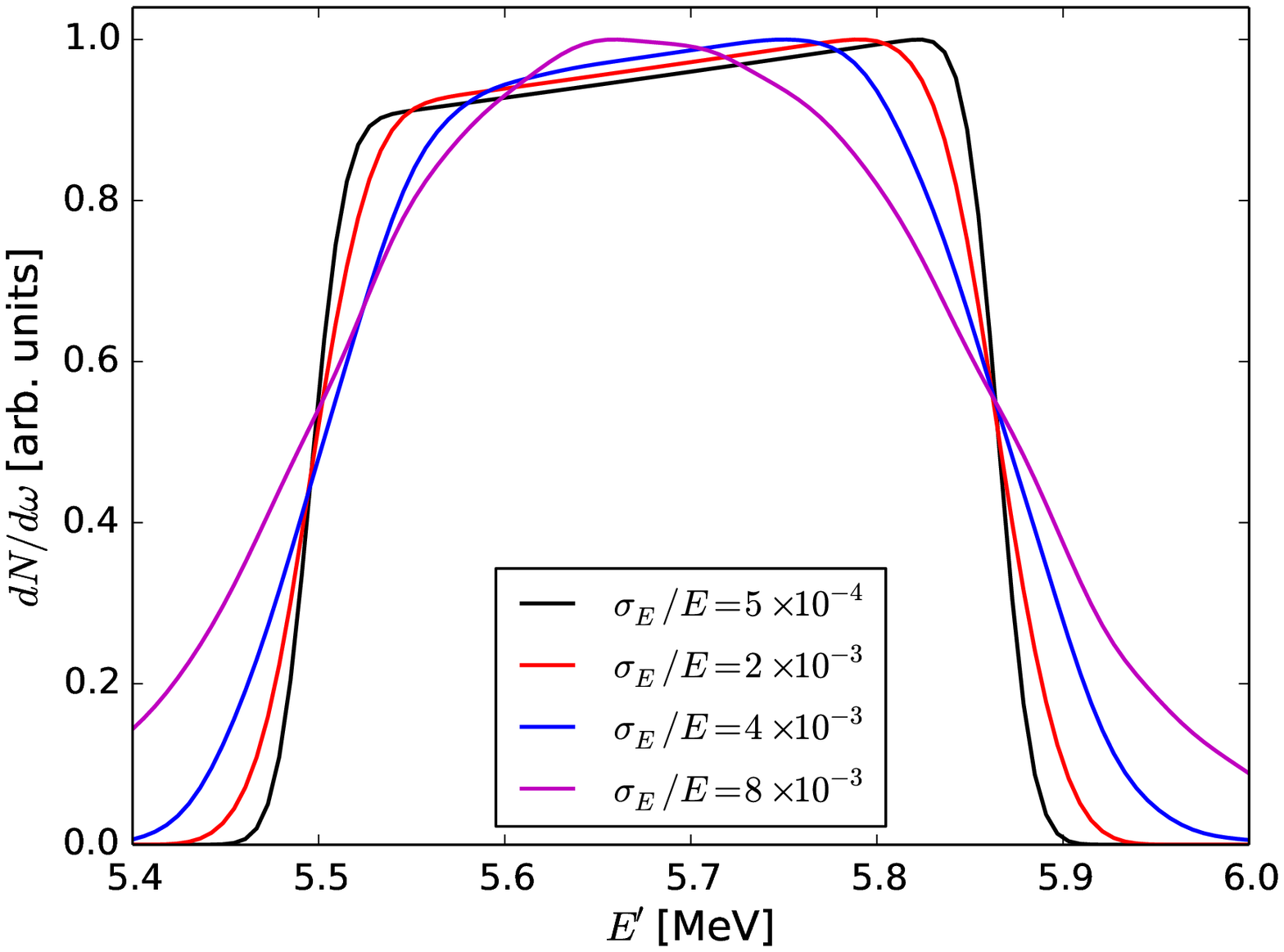}
\includegraphics[width=3.22in]{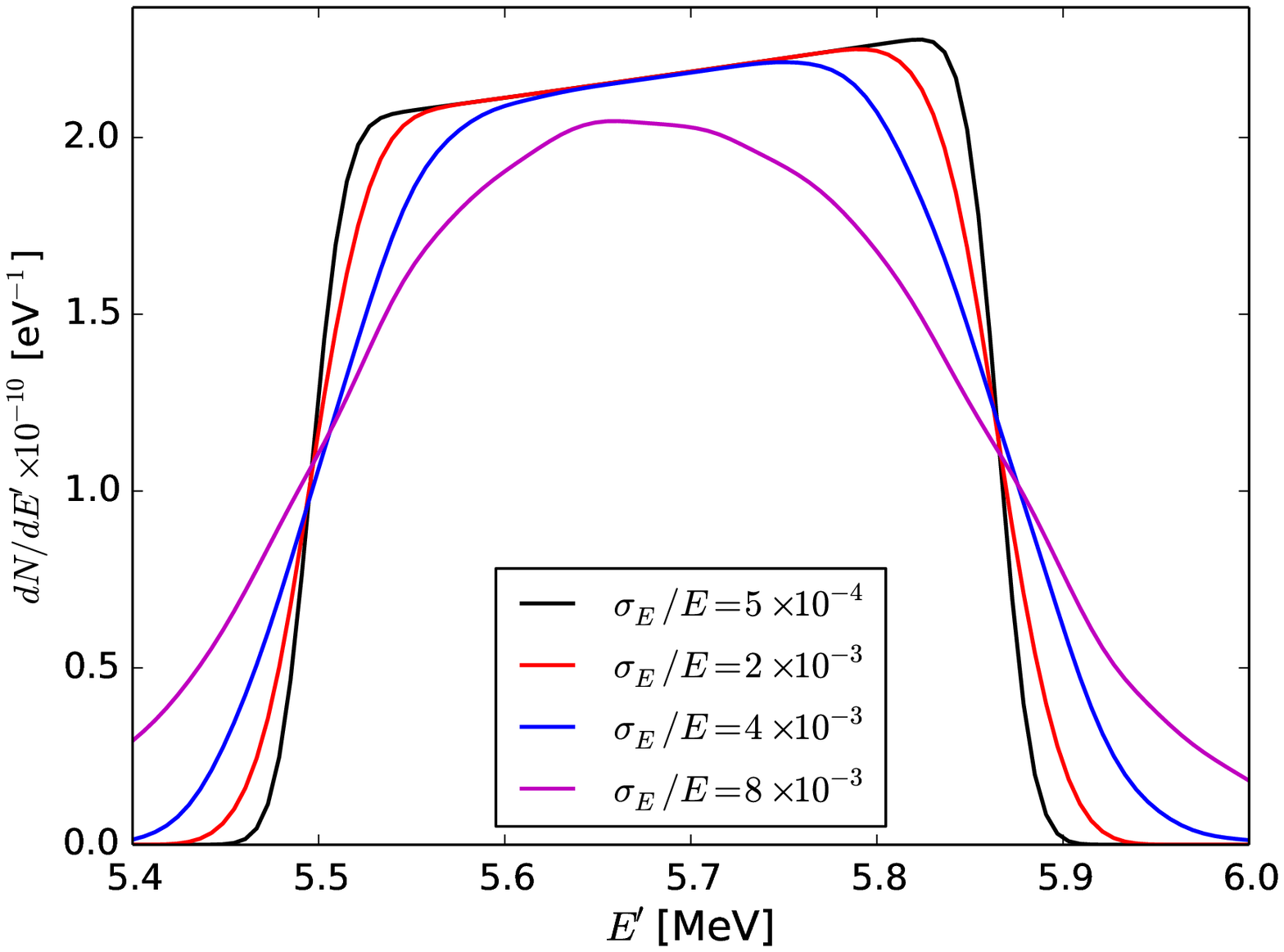}
\caption{\small{
The number density of the energy spectrum of a Compton gamma-ray beam produced
by the head-on collision of a 500 MeV electron with a 800 nm pulsed laser beam, for
different relative energy spread of the electron beam $\sigma_E$ as in Fig.~4(b) in
Ref.~\cite{slrtw09}.
The laser is collimated by an aperture with radius of 16 mm, placed 60 m downstream
from the collision point. The horizontal emittance $\epsilon_E$ is held constant at
0.05 nm rad.
Here a Gaussian laser pulse with $\sigma=50$ is used,
while the laser in Ref.~\cite{slrtw09} is a CW.
Each curve is generated by averaging 400 electrons sampling the prescribed distribution.
Left panel: Spectra scaled to their respective peak values (compare
with Fig.~4(b) of Ref.~\cite{slrtw09}). Right panel: Spectra in physical units.}}
\label{fig:fig5}
\end{center}
\end{figure}

We further check our pulsed laser model by investigating its behavior with $\sigma=50$ against other
results reported in Ref.~\cite{slrtw09} in cases when detector aperture,
emittance, and the electron beam energy spread are varied.
The dependence of the computed spectrum on detector aperture is shown
in Fig.~\ref{fig:fig3}. The left panel is in near-perfect agreement with Fig.~3(b)
of Ref.~\cite{slrtw09}; including the laser pulsing accounts for any slight differences observed. The right panel shows the non-normalized spectrum.
Figure \ref{fig:fig4} captures the dependence of the computed spectrum on
electron beam emittance. Again, the left panel is in near-perfect agreement with Fig.~4(a)
of Ref.~\cite{slrtw09} and the right panel shows the spectrum in physical units.
The dependence of the computed spectrum on electron beam energy spread is
illustrated in Fig.~\ref{fig:fig5}. The left panel is in agreement with Fig.~4(b)
of Ref.~\cite{slrtw09} and the right panel shows the non-normalized spectrum. Because we have been able to reproduce earlier results produced in an entirely different way with our code, we are highly confident in our numerical method.

\section{Scaling of Scattered Photon Energy Spread} \label{sec:LPeffects}

The effects of the energy spreads in the two colliding beam---$\sigma_{E_e}/E_e$ for
the electron beam and $\sigma_{E_p}/E_p$ for the incident photon beam---on the
linewidth of the scattered radiation have been estimated from first principles \cite{slrtw09}
as
\begin{equation} \label{Eq5}
\frac{\sigma_{E'}}{E'} \approx \sqrt{
\left(2\frac{\sigma_{E_e}}{E_e}\right)^2 + \left(\frac{\sigma_{E_p}}{E_p}\right)^2,
}
\end{equation}
where for our Gaussian model one can show
\begin{equation}
\frac{\sigma_{E_p}}{{E_p}}=\frac{1}{2\sqrt{2}\pi\sigma} \, ,
\end{equation}
However, this equation does not account for the intrinsic energy spread of the aperture
$\sigma_{a}/E_{a}$.
A more complete expression which takes this effect into consideration is
\begin{equation} \label{Eq5b}
\frac{\sigma_{E'}}{E'} \approx \sqrt{
\left(2\frac{\sigma_E}{E_e}\right)^2 + \left(\frac{\sigma_{E_p}}{E_p}\right)^2
+ \left(\frac{\sigma_{a}}{E_{a}}\right)^2
}
\end{equation}
with the aperture energy spread
\begin{equation} \label{Ea}
\frac{\sigma_{a}}{E_{a}} = \frac{\omega'_{\rm max} -\omega'_{\rm min}}{\sqrt{12}\; \omega'_{\rm mid}},
\end{equation}
and
\begin{eqnarray} \label{omegas}
\omega'_{\rm max} & = &  \omega'(\theta=0) = \frac{\omega(1+\beta)}{1-\beta+(2\hbar\omega/\gamma m c^2)} \, , \nonumber \\
\omega'_{\rm min} & = &  \omega'(\theta=\theta_a) = \frac{\omega(1+\beta)}{1-\beta\cos(\theta_a)+(\hbar\omega/\gamma m_e c^2) (1+\cos(\theta_a))} \, , \\
\omega'_{\rm mid} & = & \frac{ \left(\omega'_{\rm max} + \omega'_{\rm max}\right)}{2} \nonumber \, .
\end{eqnarray}
Equation (\ref{Ea}) quantifies the relative {\it rms} energy spread of the approximately uniform distribution of frequencies passing the aperture when $\sigma_{E_e}/E_e=0$ and $\sigma_{E_p}/E_p=0$. It follows directly from the fact that the {\it rms} width of a variable uniformly distributed between 0 and 1 is $1/\sqrt{12}$.

The energy spread due to emittance is \cite{krafftreview}
\begin{equation} \label{Ee}
\frac{\sigma_{\epsilon}}{E_{\epsilon}} = \frac{2\gamma^2\epsilon}{\beta^*},
\end{equation}
where $\beta^*$ is the electron beta function at the interaction point. Because this contribution to the spread generates an asymmetrical low energy tail and a very non-Gaussian distribution, it does not as simply combine with the other sources.

\begin{figure}
\begin{center}
\includegraphics[width=3.22in]{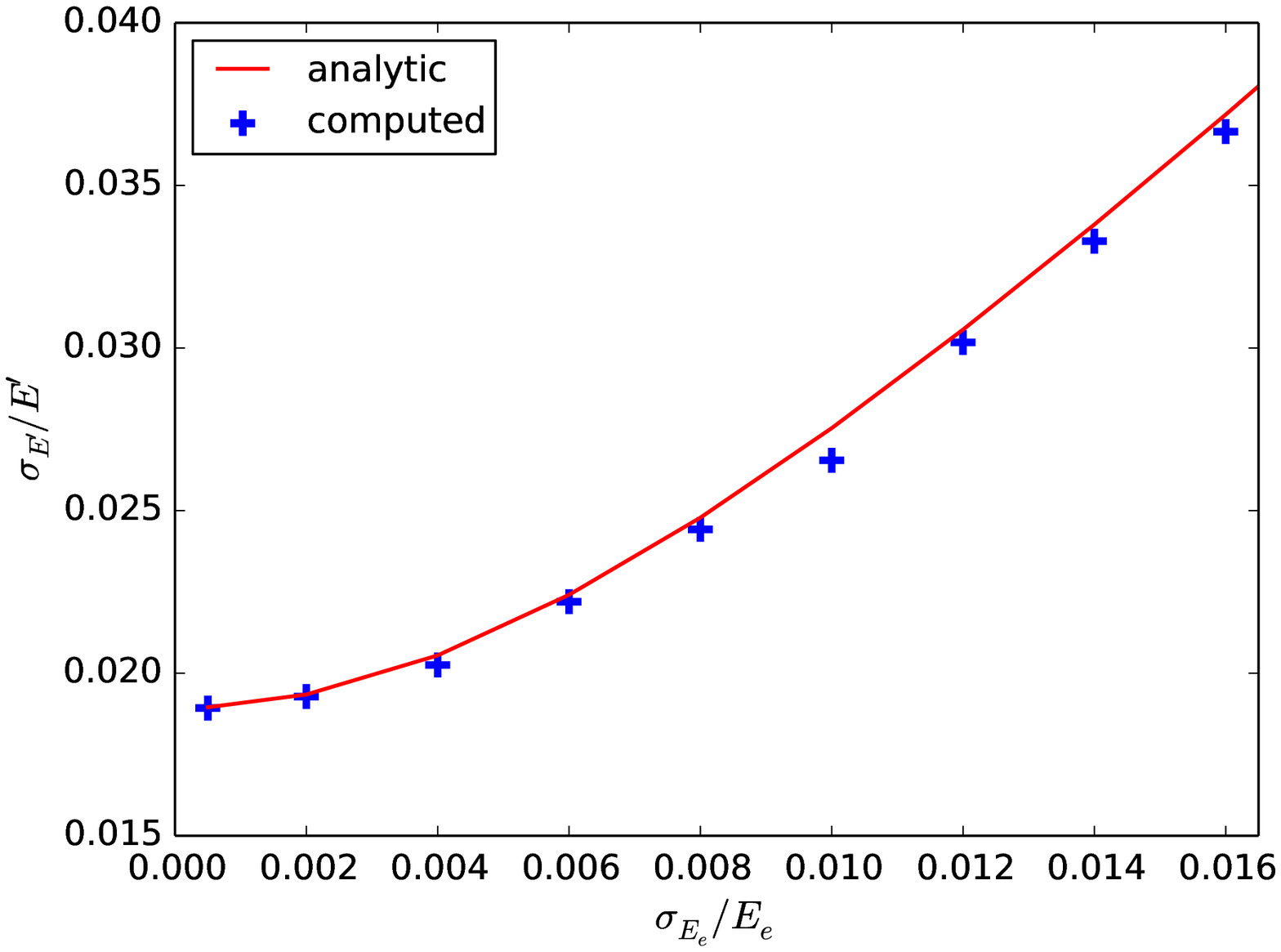}
\includegraphics[width=3.22in]{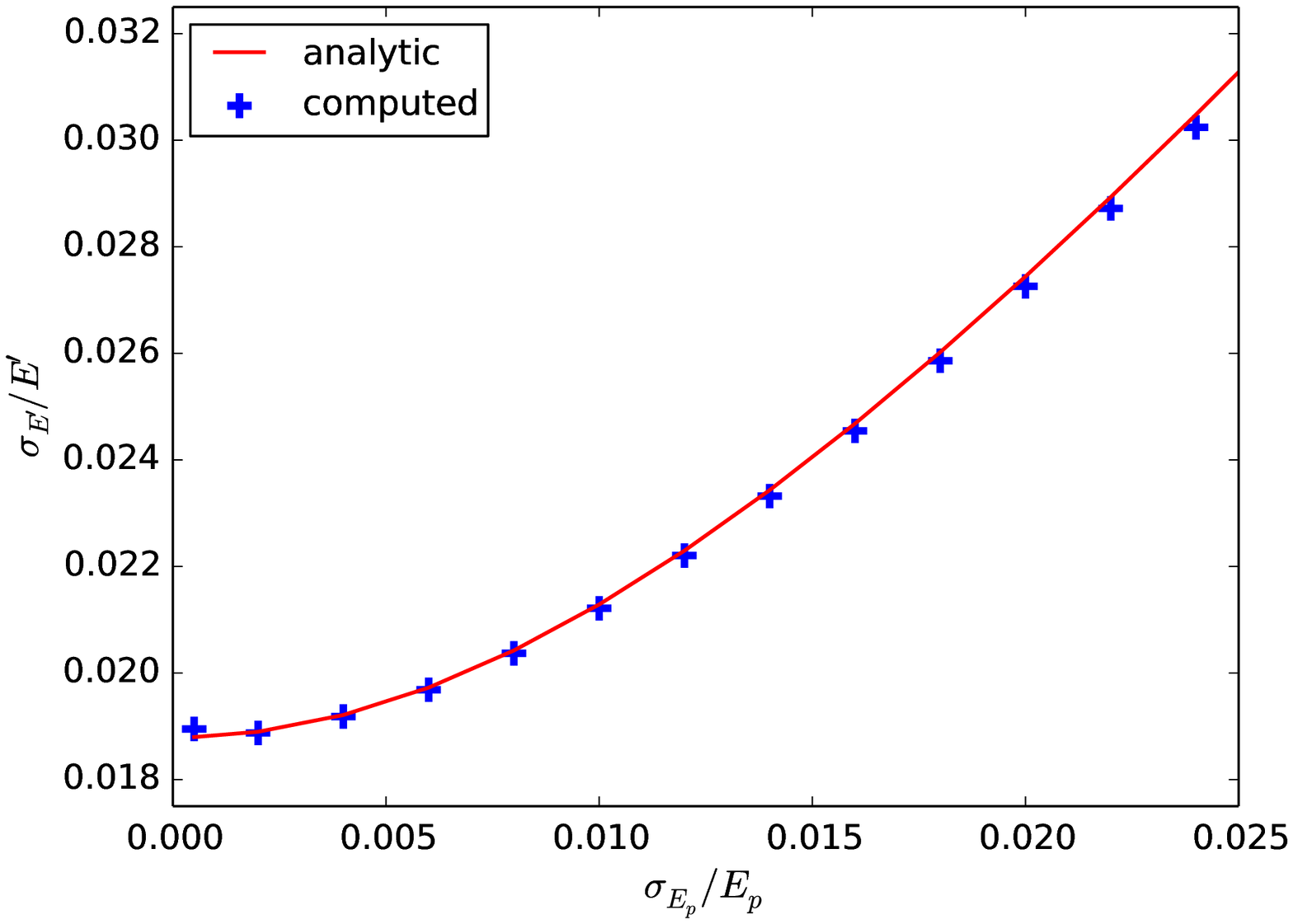}
\includegraphics[width=3.22in]{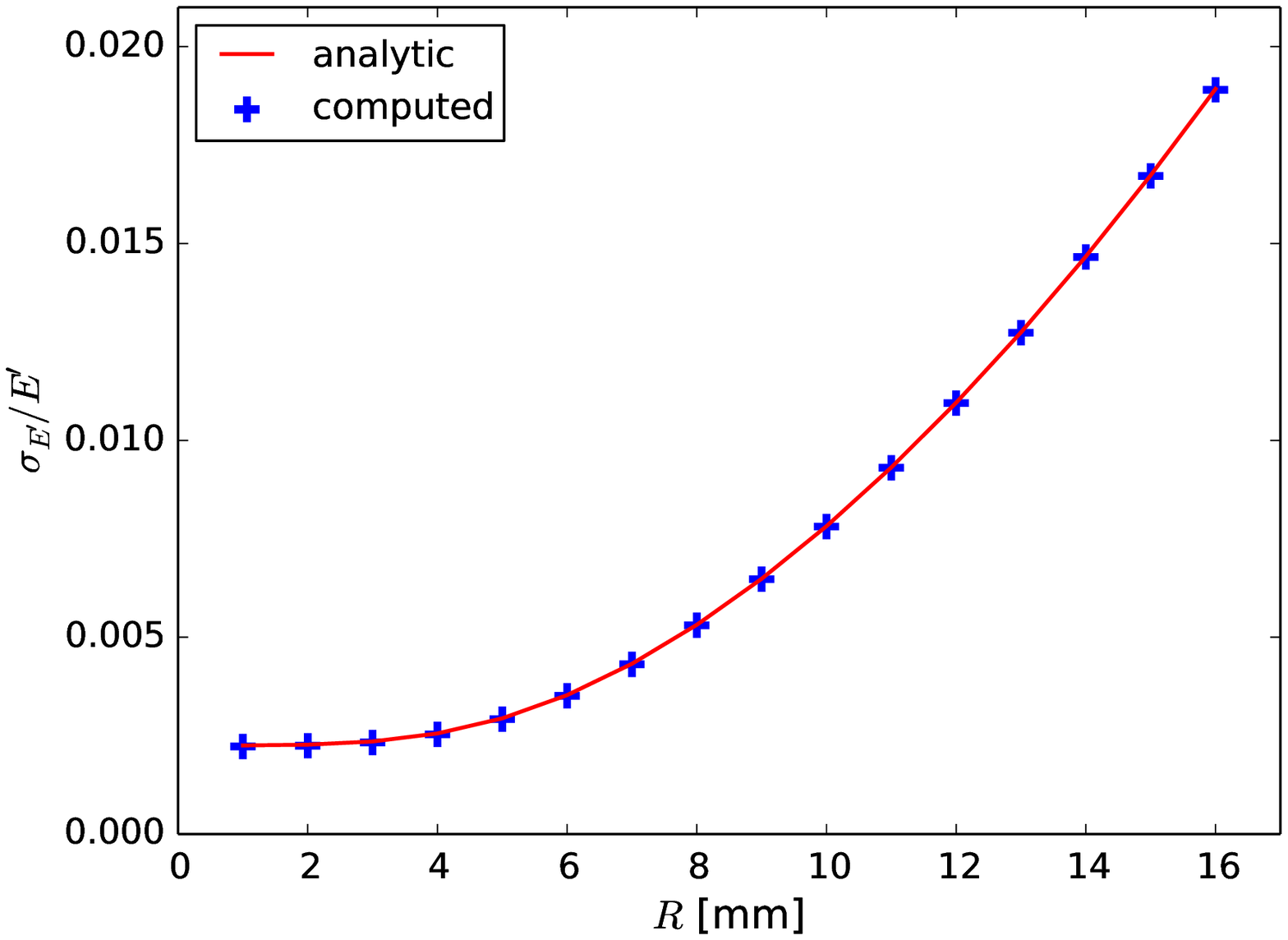}
\caption{\small{
Relationship between the energy spread of the two colliding beams---$\sigma_{E_e}/E_e$ for
the electron beam and $\sigma_{E_p}/E_p$ for the incident photon beam---and the energy
spread of the scattered radiation $\sigma_{E'}/E'$: analytically predicted from
Eq.~(\ref{Eq5b}) (red curves) and computed with our code (blue crosses).
The parameters of the simulation are 500 MeV electron beam energy with a 800 nm
pulsed laser beam, the horizontal emittance $\epsilon_E=0$. The laser is collimated by an
aperture with radius of 16 mm, placed 60 m downstream from the collision point.
Each blue point is generated by averaging 1000 electrons sampling the prescribed distribution.
Top left: Energy spread of the electron beam is varied; Gaussian laser pulse width is fixed at $\sigma=50$,
and aperture at 16 mm, placed 60 m downstream from the collision point.
Top right: Energy spread of the laser beam is varied by changing its width in physical space;
electron beam energy spread is held constant at $\sigma_{E_e}/E_e=0$ and aperture at 16 mm,
placed 60 m downstream from the collision point.
Bottom: Radius of the aperture $R$ is varied; electron beam energy spread is held constant at
$\sigma_{E_e}/E_e=0$, photon beam at $\sigma=50$.
}}
\label{fig:fig6}
\end{center}
\end{figure}

\begin{figure}
\begin{center}
\includegraphics[width=3.22in]{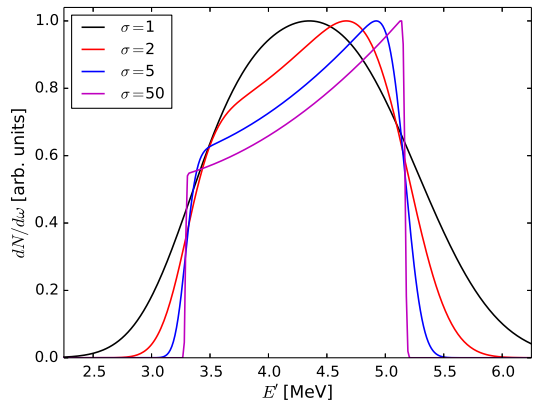}
\caption{\small{
Same as Fig.~\ref{fig:fig2}, only with varying width of the laser pulse $\sigma$.
}}
\label{fig:fig7}
\end{center}
\end{figure}

Figure \ref{fig:fig6} shows a near-perfect agreement between the above estimate and the
properties of the spectra computed with our pulsed formalism.
The effect of varying the width of the laser pulse $\sigma$ on the shape of the
backscattered radiation spectrum is illustrated in Fig.~\ref{fig:fig7}.
As the width of the laser pulse grows, the CW limit is entered, and the earlier
results of Ref.~\cite{slrtw09} apply. For short pulses (small $\sigma$), the energy spread of the
laser pulse becomes so large that it dominates the backscattered spectral linewidth.

\section{Compton Frequency Shifting} \label{sec:fshifting}

The model presented in this paper properly includes the Compton recoil of the electrons. Including this effect is vital for working with high energy, relativistic photon-electron collisions.  A series of calculations, based on parameters in two recent papers \cite{gheb,ELI1}, shows that Compton recoil is significant by comparing the full Compton calculation with that performed using the Thomson limit. The parameters are those from a recent paper by the Nebraska group \cite{gheb}, and from the new ELI - NP project in Bucharest \cite{ELI2}.

\begin{figure}
\begin{center}
\includegraphics[width=3.22in]{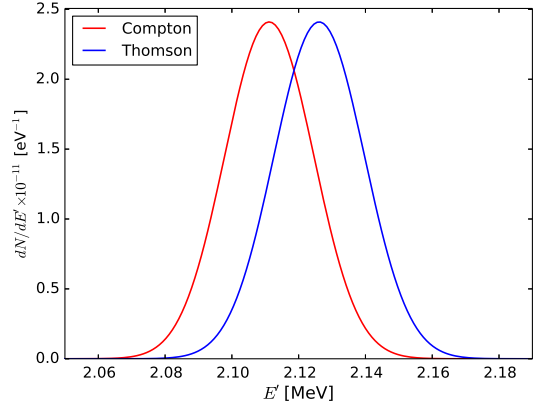}
\caption{\small{
Frequency shift between the spectra of Compton and Thomson scattering
of a single electron with $E_b = 300$ MeV, with a laser pulse with
$a_0=0.01$, $\lambda=800$ nm, $s=20.3$, captured by aperture of
$A=1/10\gamma$.
}}
\label{fig:shift}
\end{center}
\end{figure}

Figure (\ref{fig:shift}) clearly illustrates the Compton wavelength shifting.  Both spectra are computed for the parameters from Table \ref{Tab: thespar} through very small apertures, the red plot using the correct Compton computation derived in this paper and the blue plot in the Thomson limit.  For such photon-electron collisions, the electrons are well into the regime where relativistic effects are significant. Including Compton recoil decreases the scattered energy/frequency. At these electron and laser energies, the magnitude of the red shift is approximately 1\% of the scattered photon energy. Although the focus of their paper is on other issues, care should be taken in quoting the X-ray line positions given in Ref.~\cite{gheb}.

\begin{table}
\caption{Parameters used in the  backscattering spectra.  These parameters are very similar to the Ghebregziabher et al.~\cite{gheb} but the normalized vector potential has been reduced to bring the scattering event into the linear regime and eliminate the ponderomotive red shift and broadening.}
\label{Tab: thespar}
\vspace{12pt.}
\begin{center}
\begin{tabular}{lcc}\hline\hline
Parameter & Symbol & Value \\
\hline
Aperture Semi-Angle & $$ & $1/(\gamma\times\:10)$  \\
Electron Beam Energy& $E_{b}$ & 300 MeV  \\
Lorentz Factor   & $\gamma$ & 587 \\
Normalized Vector Potential& $a_{0}$ & 0.01 \\
Peak Laser Pulse Wavelength & $\lambda$ & 800 nm\\
Standard Deviation of $a(t)$ & $\sigma$ & 20.3  \\ \hline\hline
\end{tabular}
\end{center}
\end{table}

\begin{figure}
\begin{center}
\includegraphics[width=3.22in]{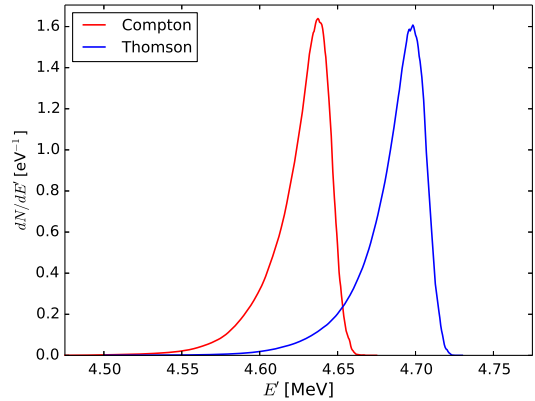}
\caption{\small{
Spectra---in both the Compton and Thomson regimes---for the ELI project, with parameters
given in Table \ref{tab:pulses}. A total of 10,000 particles were used in generating this spectrum.
}}
\label{fig:ELI}
\end{center}
\end{figure}

Frequency shifting from the recoiling electron must be properly included to predict the scattered radiation wavelength at ELI.
The properties of ELI Beam A are listed in Table \ref{tab:pulses}. We used our new approach to compute spectra for the ELI project. Figure \ref{fig:ELI} shows
spectra computed in both the Compton and Thomson regimes. For the higher energy electron beam, including recoil is clearly needed to properly account for the Compton wavelength and to obtain the correct energy in the scattered photons. Note that the Compton spectrum is different---most notably in its location in energy---from that reported in
Ref.~\cite{ELI1}. While the overall shape of the Compton spectrum is nicely reproduced in our calculation, there remains a difference in the scale which is due to an ambiguity in the definition of the aperture. Our calculations assume a full aperture $\theta_a$ of 25 $\mu$rad.

\begin{table}
    \setlength\tabcolsep{4pt}
    \caption{Main parameters of electron and laser beams for ELI project \cite{ELI2}.}
    \begin{tabular}{lll}
        \hline\hline 
        Quantity & Unit & Beam A \\
        \hline
        Charge & C & $0.25 \times 10^{-9}$ \\
        Energy & MeV & 360 \\
        Energy spread & MeV & 0.234 \\
        Normalized horizontal emittance & mm mrad & 0.65 \\
        Normalized vertical emittance & mm mrad & 0.6 \\
        Laser wavelength & $\mu$m & 0.523 \\
        Laser energy & J & 1 \\
        Laser rms time duration & ps & 4 \\
        Laser waist & $\mu$m & 35 \\
       \hline\hline
    \end{tabular} \label{tab:pulses}
\end{table}


\section{Proposed ODU Compton Source} \label{sec:ODUsource}

Superconducting RF linacs provide a means to a high average brilliance compact source of up to 12 keV X-rays. The ODU design is built on a pioneering vision developed in collaboration with scientists at MIT \cite{graves}. At present, the design has been developed to the point where full front-to-end simulations of the accelerator performance exist. The results of these simulations can be used to make predictions of the energy spectrum produced in an inverse Compton source, and to help further optimize the source design by providing feedback on those elements of the design most important for achieving high brilliance.

\subsection{Design Elements}

\begin{figure}
\begin{center}
\includegraphics[width=3.22in]{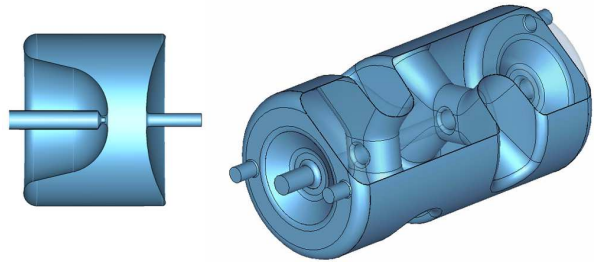}
\caption{\small{
A cross section picture of the SRF gun (left) and the SRF double-spoke accelerating cavity (right).
}}
\label{fig:gund}
\end{center}
\end{figure}

The ODU design consists of an accelerating section, operated at 500 MHz and 4.2 K, followed by a final focusing section comprised of three quadrupoles. The accelerating section begins with a re-entrant SRF gun, followed by four double-spoke SRF cavities. These two structures are shown in Fig.~\ref{fig:gund}  \cite{IPAC2015}.

The concept for the SRF gun was introduced over 20 years ago \cite{michalkethesis}. In the last ten years, the Naval Postgraduate School, Brookhaven National Lab, and University of Wisconsin have commissioned re-entrant SRF guns which operate at 4.2 K \cite{Arnold}. For the ODU design, it was needed to produce a bunch with ultra-low emittance, and the gun geometry was altered accordingly. The geometry
was mainly altered around the nose-cone containing the cathode assembly, resulting in radial electric fields within
the gun. These fields produce focusing of the bunch, making a solenoid for emittance compensation superfluous \cite{IPAC2015}.

\begin{figure}
\begin{center}
\includegraphics[width=3.22in]{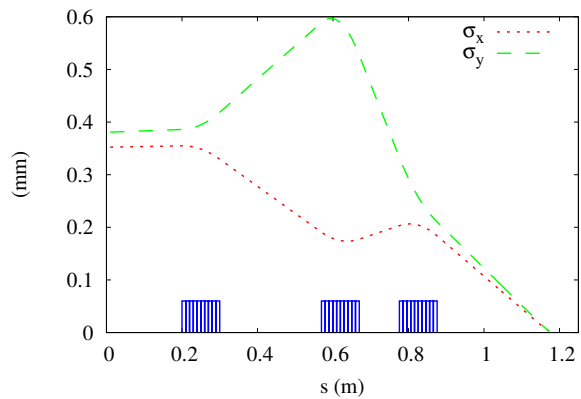}
\caption{\small{
The transverse spot size of the beam as it is focussed down into a small spot size.
}}
\label{fig:shp}
\end{center}
\end{figure}

Until recently, accelerating electrons near the speed of light has not been attempted with multi-spoke
cavities, largely because of the well-established and successful performance of TM-type cavities.
However, multi-spoke cavities are familiar options for accelerating ions. Previous studies of multi-spoke
cavities suggest strongly that they are a viable option for accelerating electrons \cite{sato,wp2}, and they provide a path to operating at 4.2 K through a low-frequency accelerator that is reasonably compact. The double-spoke cavities comprising the linear accelerator (linac) were designed by Christopher Hopper in an ODU dissertation \cite{chris,PRABacc} and developed and tested in collaboration with Jefferson Lab \cite{DoubleSpoke,park1}. The bunch exiting the linac passes through three quadrupoles. Figure \ref{fig:shp} shows the horizontal and vertical size of the bunch as it traverses the quadrupoles, before it is focused down to a small spot size.  Table \ref{tab:FINAL_IP} lists the properties of the bunch at the collision point.

\begin{table}
\caption{Properties of electron bunch at the collision point. }
\begin{center}
\begin{tabular}{lcc}
\hline\hline
Parameter & Quantity & Units \\
\hline
kinetic energy & 25.0 & MeV \\
bunch charge & 10.0 & pC \\
\textit{rms} energy spread & 3.44 & keV \\
$\epsilon^N_{x,\mathrm{rms}}$ & 0.10 & mm-mrad \\
$\epsilon^N_{y,\mathrm{rms}}$ & 0.13 & mm-mrad \\
$\sigma_x$ & 3.4 & $\mu$m \\
$\sigma_y$ & 3.8 & $\mu$m \\
$\beta_x$ & 5.4 & mm \\
$\beta_y$ & 5.4 & mm \\
FWHM bunch length & 3 & psec \\
$\sigma_z$ & 0.58 & mm \\
\hline\hline
\end{tabular} \label{tab:FINAL_IP}
\end{center}
\end{table}

\subsection{Tracking to Collision}

\begin{figure}
\begin{center}
\includegraphics[width=3.22in]{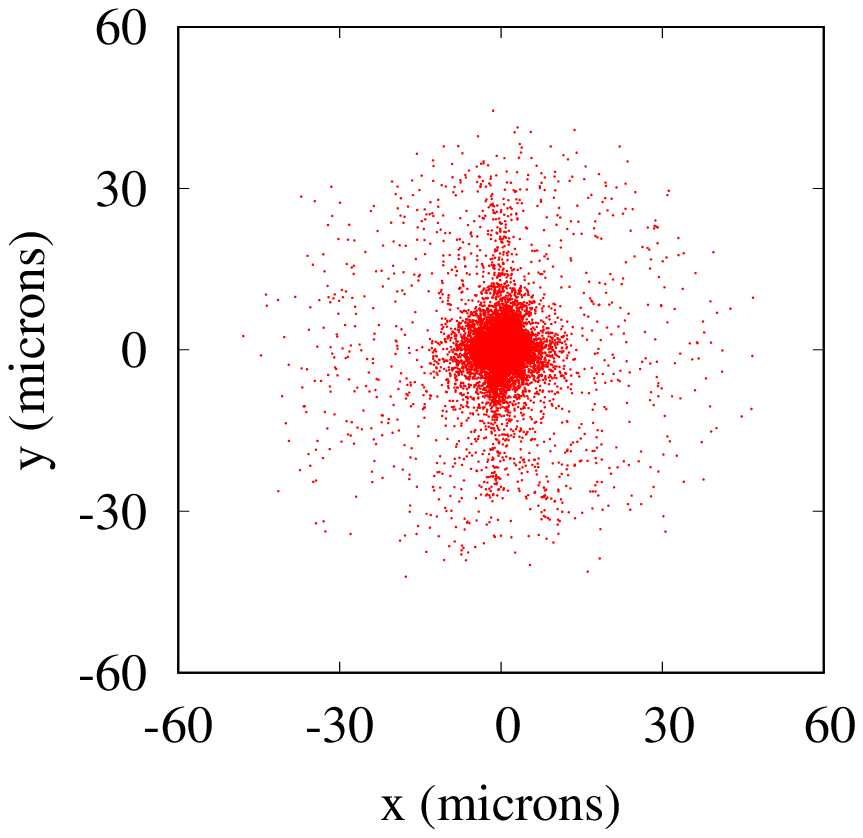}
\includegraphics[width=3.22in]{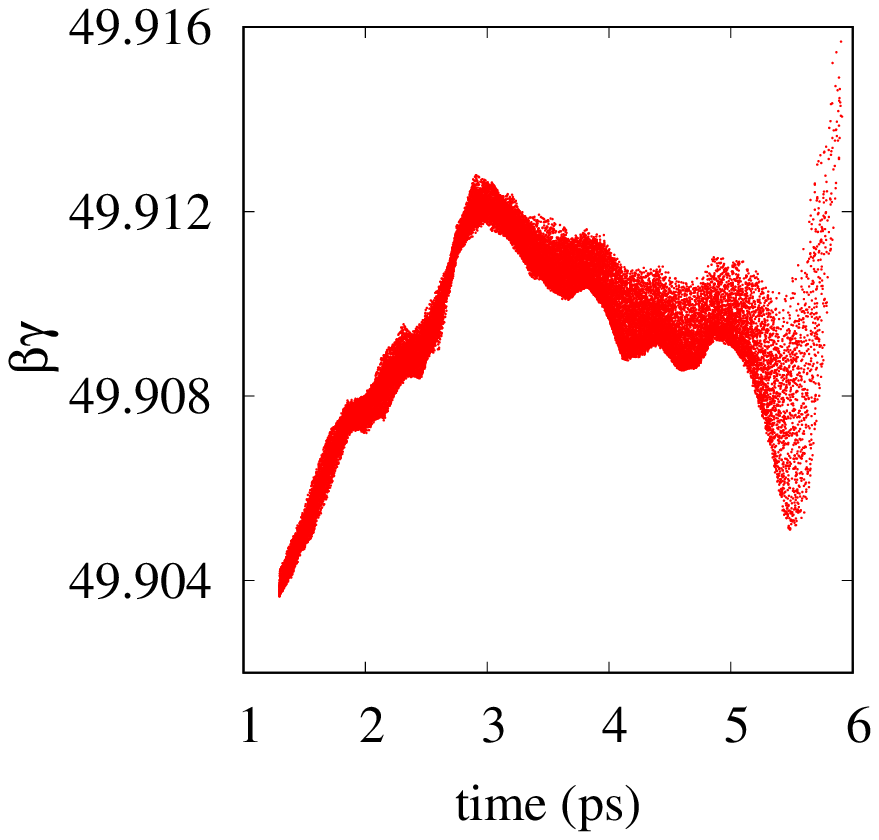}
\includegraphics[width=3.22in]{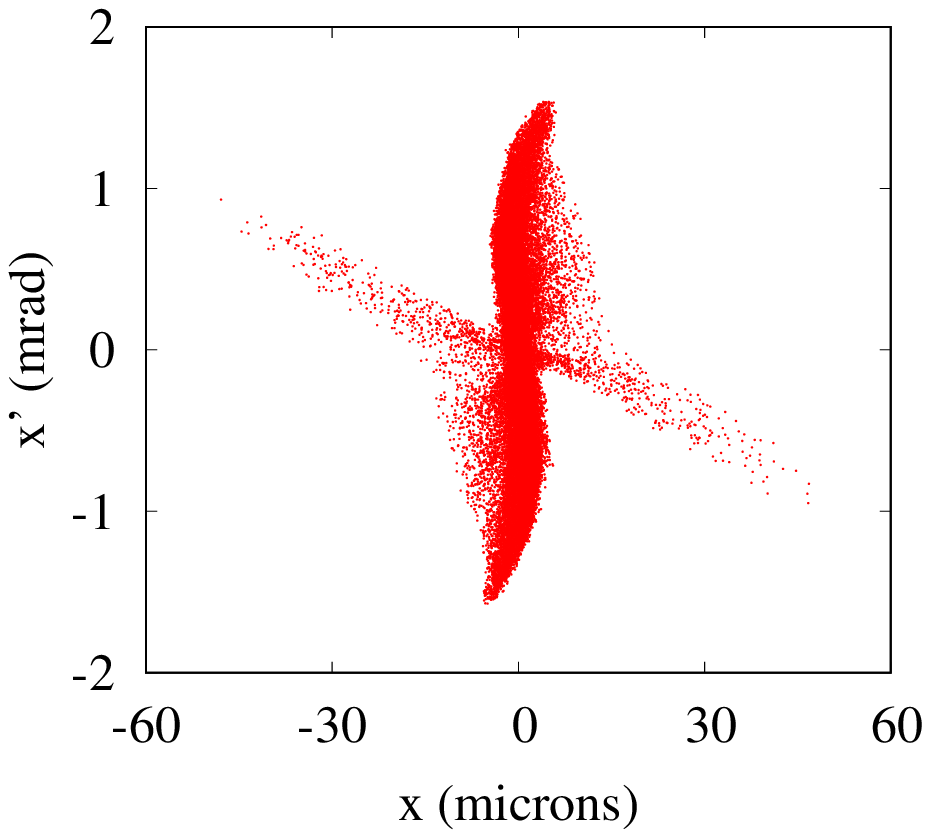}
\includegraphics[width=3.22in]{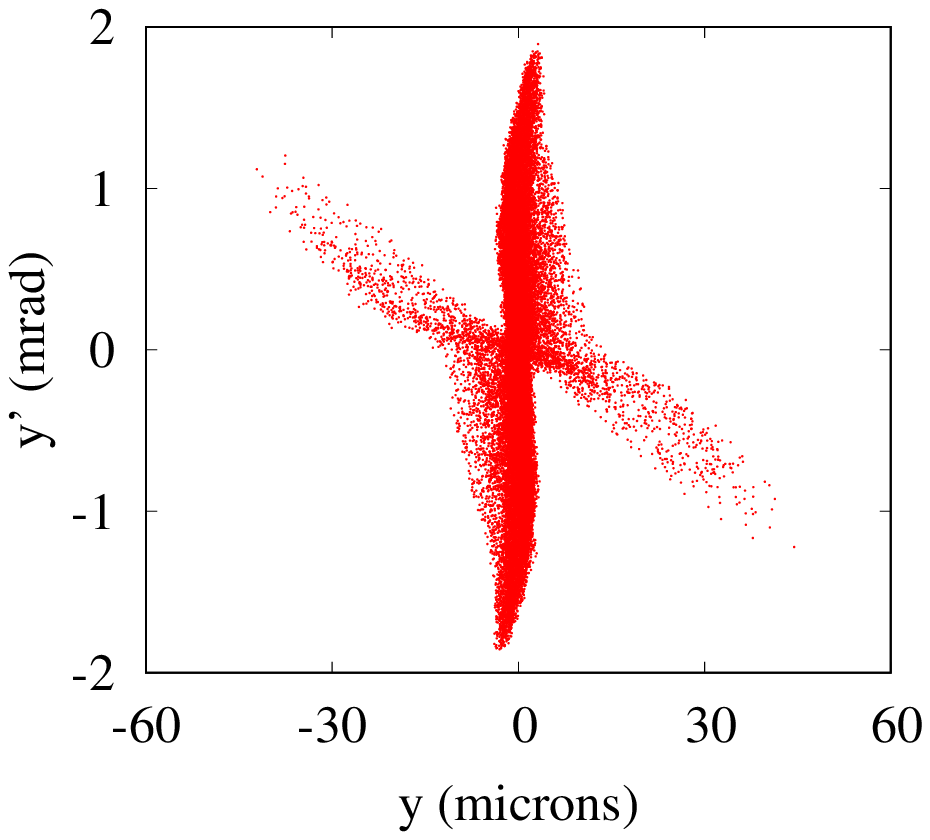}
\caption{\small{The beam spot (upper left), longitudinal phase space (upper right), horizontal phase space (lower left),
and vertical phase space (lower right) of the electron bunch at
the collision point.}}
\label{fig:ipcol}
\end{center}
\end{figure}

The electromagnetic field modes of the SRF gun and the SRF double-spoke cavity are calculated by Superfish and CST MICROWAVE STUDIO (CST MWS) respectively \cite{superf,CSTMS}. Utilizing these calculated electromagnetic fields, IMPACT-T tracked a defined particle bunch off the cathode and through the accelerating linac \cite{ImpactT}.
Afterwards, tools were used to translate the coordinates of the electrons in the bunch into the SDDS format {\tt elegant} \cite{elegant} requires, and {\tt elegant} tracks the bunch as it traverses the three quadrupoles that comprise the final focusing section.
Figure~\ref{fig:ipcol} shows simulation calculations of the beam spot and the longitudinal, horizontal, and vertical phase spaces at the collision point.

\subsection{X-ray Yield}

\begin{figure}
\begin{center}
\includegraphics[width=3.22in]{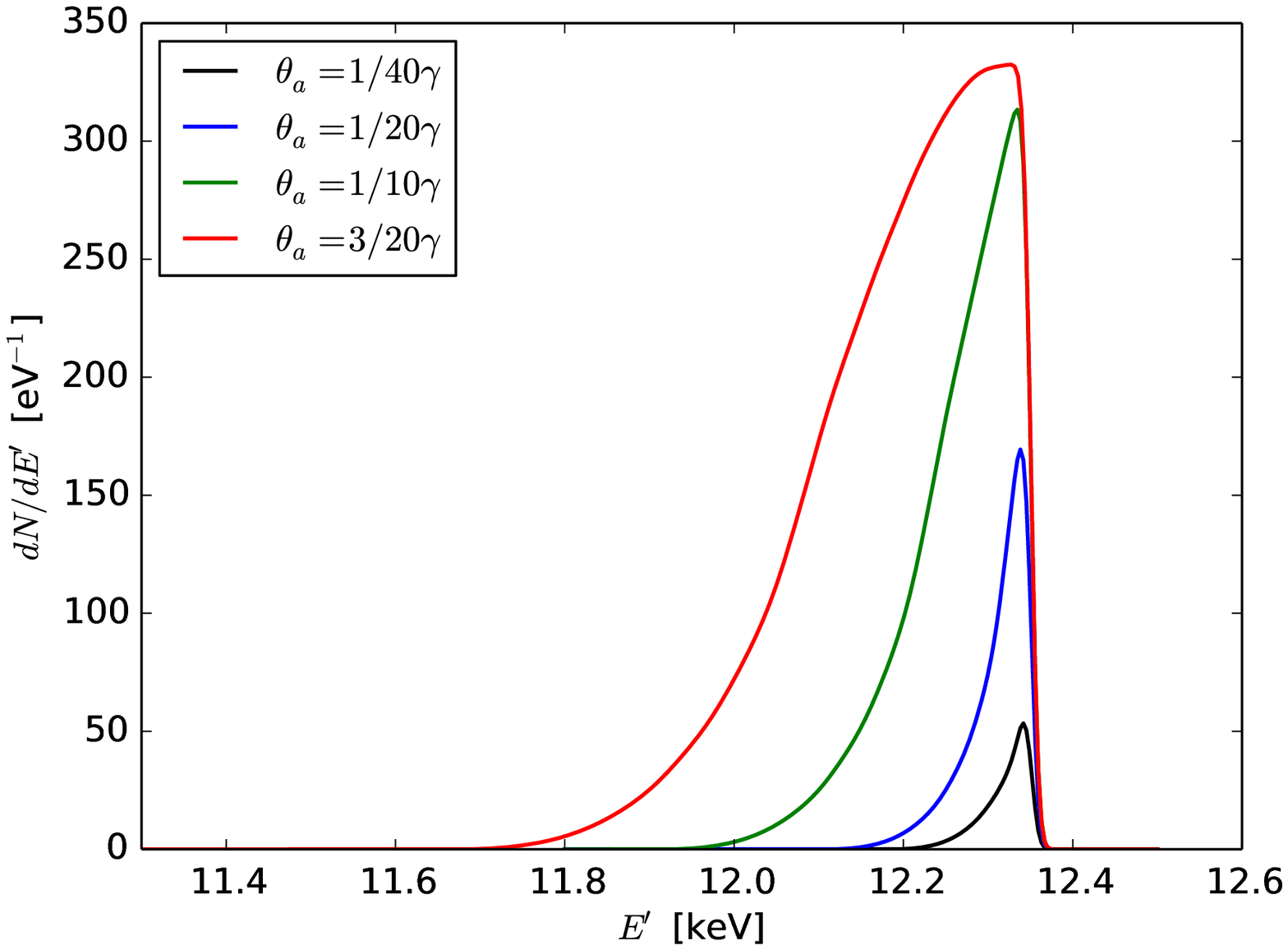}
\includegraphics[width=3.22in]{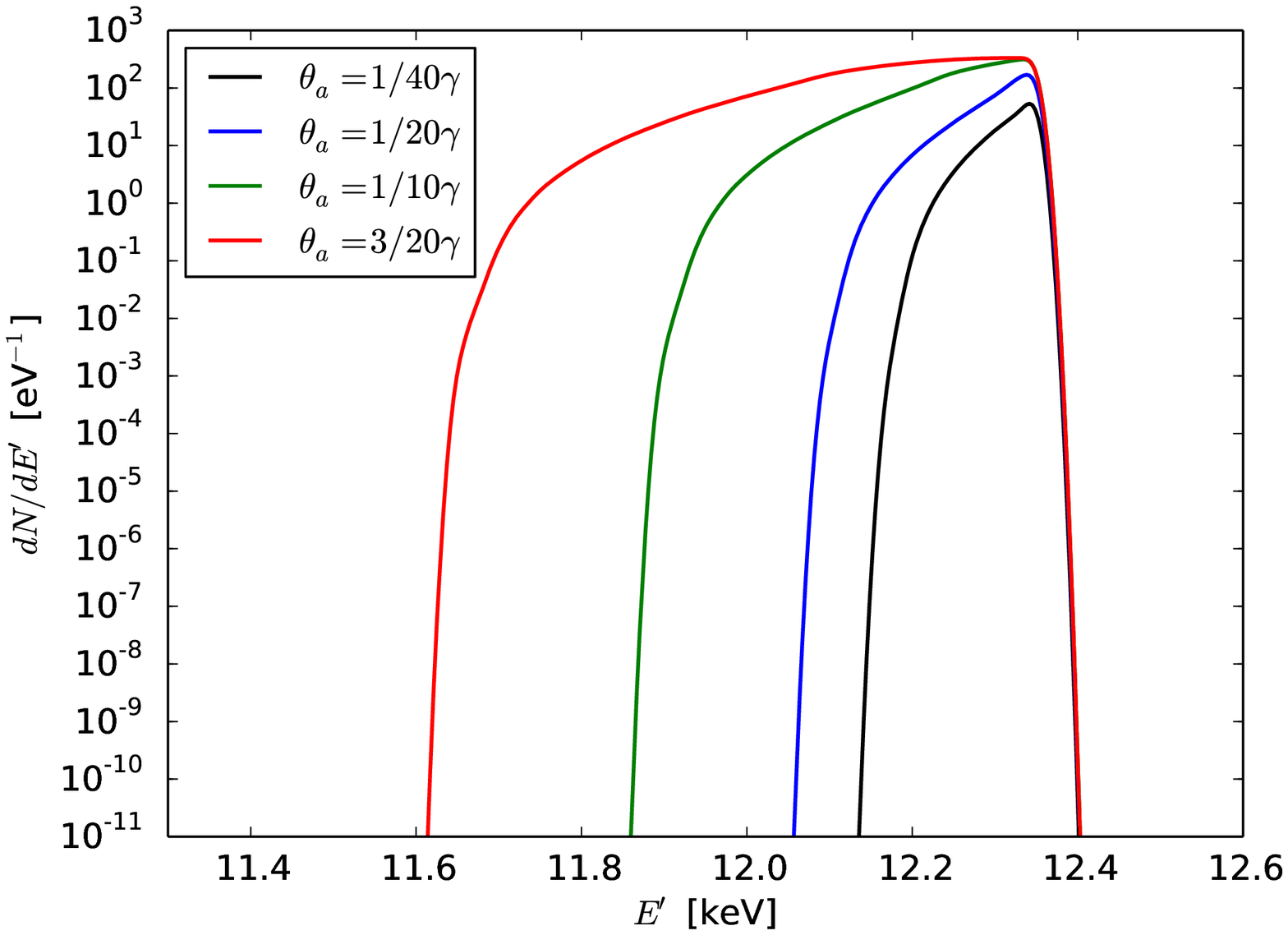}
\caption{\small{
Number spectra for the Old Dominion University Compton source with 10 pC electron bunch charge.
Left: For apertures $1/40\gamma$, $1/20\gamma$, and $3/20\gamma$, 4,000 particles were used in generating each curve. For aperture $1/10\gamma$, 48,756 particles were used in generating the plot.
Right: The same as the panel on the left, except on the log scale.
}}
\label{fig:ODUspect}
\end{center}
\end{figure}

The simulations were used to generate the beam distribution at collision as represented by 48,756 simulation particles. This distribution was then used to determine the scattered photon spectrum through various apertures. The resulting spectra are shown in Fig.~\ref{fig:ODUspect}, generated by colliding 4,000 particles from the
ensemble of 48,756 tracked. The right panel of the figure shows the same spectra
on the log scale, demonstrating that the accuracy of this calculation method allows one to evaluate the importance of the tails of the electron distribution on the final result. In addition, the radiation spectrum was calculated once with the full complement of electrons at the aperture of $1/10\gamma$, yielding negligible difference with the 4,000 particle calculation shown. With assumptions about the scattering laser, it is possible
to determine the X-ray source that a head-on collision of these two beams will provide. We assumed a 1 MW
circulating power laser, with a spot size of 3.2 $\mu$m and a wavelength of 1 $\mu$m \cite{sato}.

Consistent with the very small transverse source size, the average brilliance of the photon beam is obtained from a pin-hole measurement
\begin{equation}
B=\lim_{\theta_a\rightarrow 0}\frac{F}{2\pi\sigma_x\sigma_y\pi \theta_a^2},
\end{equation}
where $F$ is the number of photons in a 0.1$\%$ bandwidth transmitted through the aperture. Collecting the results from the $1/40\gamma$ figure and remembering 0.1$\%$ of 12 keV is 12 eV, the maximum number of photons in a 0.1$\%$ bandwidth through the aperture is 600 at 10 pC.
The average flux and brilliance for 6.242$\times10^{15}$ electrons per second (10 pC at 100 MHz) is shown in Table \ref{tab:xrprops}. Essentially because of the small spot size in collision and the high repetition rate of collisions, this result is world-leading for Compton sources
\cite{deitrickthes}.

\begin{table}
\caption{X-ray source properties.}
\label{TB:XRAY}
\begin{center}
\begin{tabular}{lcc}
\hline\hline
Parameter & Quantity & Units \\
\hline
$N_{\gamma}$ & 1.4 $\times 10^6$ & photons \\
Flux & 1.4 $\times 10^{14}$ & ph/sec \\
Full Flux in 0.1\%BW & 2.1 $\times 10^{11}$ & ph/(s-0.1\%BW) \\
Average brilliance & 1.0 $\times 10^{15}$ & ph/(s-mm$^2$-mrad$^2$-0.1\%BW) \\
\hline\hline
\end{tabular}
\label{tab:xrprops}
\end{center}
\end{table}

\section{On the Circular and Elliptically Polarized Cases}
\label{sec:circular}

In the course of this work a question arose about the correct generalization of the Klein-Nishina formula for scattering of circularly polarized photons. After reviewing relevant literature, some of which was contradictory or incorrect, a proper calculation was completed which is documented in this section. In particular, the calculation reduces to the correct beam-frame results given by Stedman and Pooke \cite{Steadpooke}, but covers general kinematics as in the linearly polarized case above. Our concern is with scattering of polarized photons from unpolarized electrons. Others, concerned with electron polarimetry, have written out solutions for scattering from polarized electrons.

The beam frame Klein-Nishina scattering differential cross section is sometimes presented as \cite{jar,feyn,jack2nd} [cf. Eq.~(\ref{klnilincros})], 
\begin{equation} \label{wrgone}
\frac{{\mathrm{d}\sigma }}{{\mathrm{d}\Omega _b }} 
= \frac{{r_e^2 }}{4}\left(
{\frac{{\omega '_b }}{{\omega _b }}} \right)^2 \left[ {\frac{{\omega '_b }}{{\omega _b }} + \frac{{\omega _b }}{{\omega '_b }} - 2 + 4\big| {\varepsilon _b  \cdot \varepsilon '^*_b } \big|^2 } \right],
\end{equation}
The fact that circular polarization is discussed elsewhere within all of these references might lead one to conclude that the formula is valid for general complex polarization vectors, e.g., for scattering of elliptically or circularly polarized lasers. This conclusion is incorrect; Eq.~(\ref{wrgone}) has validity for linear polarization only because then the scalar product involves purely real polarization vectors and $\left| {\varepsilon _b  \cdot \varepsilon '^*_b } \right|^2 = \left( {\varepsilon _b  \cdot \varepsilon '_b } \right)^2 $, as above. However, the differential cross section in Eq.~(\ref{wrgone}) is not valid for more general complex cases.

	The proper beam frame differential cross section has been provided by Stedman and Pooke \cite{Steadpooke}
\begin{equation} \label{stpk}
\frac{{\mathrm{d}\sigma }}{{\mathrm{d}\Omega _b }}
 = \frac{r_e^2 }{4}\left(\frac{\omega '_b }{\omega _b }\right)^2 \left[ {\left( \frac{\omega '_b }{\omega _b } + \frac{\omega _b }{\omega '_b } \right)\left( {1 - \big| {\varepsilon _b  \cdot \varepsilon '_b } \big|^2  + \big| {\varepsilon _b  \cdot \varepsilon '^*_b } \big|^2 } \right) + 2\left( {\big| {\varepsilon _b  \cdot \varepsilon '_b } \big|^2  + \big| {\varepsilon _b  \cdot \varepsilon '^*_b } \big|^2  - 1} \right)} \right].
\end{equation}
In addition, these authors point to the incorrect assumption leading to the derivation of Eq.~(\ref{wrgone}) when complex polarization vectors are involved: $(\varepsilon_{b\mu}\gamma^\mu)(\varepsilon^*_{b\nu}\gamma^\nu)\ne-1$ where the multiplication here is between the summed Dirac matrices. It has been verified that when the correct anti-commutator relation $[\varepsilon_{b\mu}\gamma^\mu,\varepsilon^*_{b\nu}\gamma^\nu]_+=2\varepsilon_b\cdot\varepsilon^*_b$ is used when performing the traces to evaluate the differential cross section, Eq.~(\ref{stpk}) results. Clearly, for linear polarization $\varepsilon_b=\varepsilon^*_b$ and $\varepsilon'_b=\varepsilon'^*_b$, and Eq.~(\ref{stpk}) leads directly to Eq.~(\ref{klnilincros}). Also, Eq.~(\ref{stpk}) is equivalent to the corrected expression
\begin{equation}
\frac{{\mathrm{d}\sigma }}{{\mathrm{d}\Omega _b }}
 = r_e^2 \left( {\frac{{\omega '_b }}{{\omega _b }}} \right)^2 \left[ \big| \bm{\varepsilon} _b  \cdot \bm{\varepsilon} '^*_b \big|^2  + \frac{{\left( {\omega _b  - \omega '_b } \right)^2 }}{{4\omega _b \omega '_b }}{\Big( 1 + \left( \bm{\varepsilon}_b  \times \bm{\varepsilon} ^*_b \right) \cdot \left( \bm{\varepsilon} '^*_b  \times \bm{ \varepsilon} '_b  \right) \Big)} \right],
\end{equation}
which appears in the third edition of Jackson's text \cite{jack3rd}.

Clearly, the procedure followed in the linear polarization case translates here. The lab frame cross section may be written down by inspection
\begin{equation}
\begin{array}{rl}
\dfrac{{\mathrm{d}\sigma }}{{\mathrm{d}\Omega }} 
=& \dfrac{r_e^2 }{4\gamma ^2(1 - \bm{\beta}  \cdot \bm{\hat k})^2 }\left(\dfrac{\omega '}{\omega} \right)^2 
\left[ 
 \left( 
\dfrac{\omega '(1 - \bm{\beta}  \cdot \bm{\hat k}')}
{\omega (1 - \bm{\beta}  \cdot \bm{\hat k})} 
 + \dfrac{\omega (1 - \bm{\beta}  \cdot \bm{\hat k})}
{\omega '(1 - \bm{\beta}  \cdot \bm{\hat k}')} 
\right)
\left(1 - \big| P(\varepsilon ,\varepsilon ') \big|^2 
+ \big|P(\varepsilon ,\varepsilon '^* ) \big|^2  \right)
\right. \\
 &\left. \hspace{4cm} +\ 2\ \left( \big| P(\varepsilon ,\varepsilon ') \big|^2  + \big|P( \varepsilon ,\varepsilon '^* ) \big|^2  - 1 \right) \right] .
 \end{array}
\end{equation}
This equation extends Eq.~(\ref{crosssect1}) to include cases with arbitrary complex polarization and agrees with Eq.~(3) of Ref.~\cite{bocaf1} by applying energy-momentum conservation to eliminate $p_f$.
In sources where a circularly polarized laser beam is scattered, but the final polarization is not observed, the polarization sum is modified.
For general complex polarization vectors Eq.~(\ref{polsum}) becomes
\begin{equation} \label{othersum}
\begin{array}{rl}
- P^\mu(\varepsilon) P_\mu(\varepsilon^*) 
&=- P^\mu(\varepsilon) P_\mu^*(\varepsilon)\ =\ 
P\left( {\varepsilon ,\varepsilon '_1 } \right)P\left( {\varepsilon ^* ,\varepsilon '^*_1 } \right) + P\left( {\varepsilon ,\varepsilon '_2 } \right)P\left( {\varepsilon ^* ,\varepsilon '^*_2 } \right) \\

 = 1 &-\ m^2 c^2 \left[ 
\dfrac{{(k' \cdot \varepsilon)(k' \cdot \varepsilon ^* )}}{{( p_i  \cdot k')^2 }} - \dfrac{{(p_i  \cdot \varepsilon )(k' \cdot \varepsilon ^* ) + (k' \cdot \varepsilon )(p_i  \cdot \varepsilon ^* )}}{{(p_i  \cdot k)(p_i  \cdot k')^2 }}(k \cdot k') 
+ \dfrac{{(p_i  \cdot \varepsilon )(p_i  \cdot \varepsilon ^* )}}{{(p_i  \cdot k)^2 (p_i  \cdot k')^2 }}(k \cdot k')^2 \right] \\
 \end{array}
\end{equation}
now for any two orthonormal complex polarization vectors $\varepsilon'_1$  and $\varepsilon'_2$ orthogonal to the propagation vector  $k'$.  Because Eq.~(\ref{othersum}) is identical under the interchange $\varepsilon\leftrightarrow\varepsilon^*$, $P(\varepsilon,\varepsilon'^*_1)
P(\varepsilon^*,\varepsilon'_1)+P(\varepsilon,\varepsilon'^*_2)
P(\varepsilon^*,\varepsilon'_2)$ evaluates identically, and the summed differential cross section is
\begin{equation} \label{crosscirc}
\begin{array}{rl}
\dfrac{{\mathrm{d}\sigma }}{{\mathrm{d}\Omega }} =& \dfrac{{r_e^2 }}{{2\gamma ^2 (1 - \bm{\beta}  \cdot \bm{\hat k})^2 }}
\left(\dfrac{\omega '}{\omega } \right)^2
 \left[
\dfrac{{\omega '(1 - \bm{\beta}  \cdot \bm{\hat k}')}}
{{\omega(1 - \bm{\beta}  \cdot \bm{\hat k})}}
+ \dfrac{{\omega(1 - \bm{\beta}  \cdot \bm{\hat k})}}
{{\omega '(1 - \bm{\beta}  \cdot \bm{\hat k}')}} \right. \\

&\left. \hspace{0.25cm} -2m^2 c^2 \left( \dfrac{{( k' \cdot \varepsilon )(k' \cdot \varepsilon ^* )}}{{(p_i  \cdot k')^2 }}
  - \dfrac{{(p_i  \cdot \varepsilon )(k' \cdot \varepsilon ^* ) + (k' \cdot \varepsilon )(p_i  \cdot \varepsilon ^* )}}
{{(p_i  \cdot k)(p_i  \cdot k')^2 }}(k \cdot k')
 + \dfrac{{(p_i  \cdot \varepsilon )(p_i  \cdot \varepsilon ^* )}}
{{(p_i  \cdot k)^2 (p_i  \cdot k')^2 }}(k \cdot k')^2 \right)  \right] . \\
 \end{array}
\end{equation}

It should be emphasized that this result is correct for circular polarization vectors and reduces to Eq.~(\ref{cross2}) for linear polarization. Equation (\ref{crosscirc}) may be averaged over the initial spin by effecting initial polarization summations. The correct electron and photon spin-averaged differential cross section emerges \cite{pessch,landp}.

\section{Summary} \label{sec:summary}

In this paper a novel calculation prescription is used to
determine the emission characteristics of the scattered radiation
in a Compton back-scatter source. The model we have developed has been exercised to precisely calculate the photon energy distributions from Compton scattering events. The calculations are quite general, incorporating beam emittance, beam energy spread, laser photon spread, and the full Compton effect. The final form of the scattering distribution is quite convenient for computer implementation and simulation with computer-calculated electron beam distributions.

The calculations sum the full electron rest frame Klein-Nishina scattering cross section, suitably transformed to the lab frame, on an electron by electron basis. Although somewhat ``brute-force" and moderately computationally expensive, such a calculational approach has several advantages. Firstly, the model accurately accounts for the details in the spectra that are generated from, e.g., non-Gaussian particle distributions or other complicated particle phase spaces. Secondly, it is straightforward to incorporate into the model pulsed incident lasers in the plane-wave approximation. And thirdly, and most significantly, the model is simple and straightforward to implement computationally. Any numerical problems we have observed in executing our computations have been due to causes easily understood and straightforwardly addressed.

As test and benchmarking cases, we have confirmed the results of the Duke group \cite{slrtw09} and reconsidered a calculation of the Nebraska group \cite{gheb}, and corrected and confirmed a calculation made for ELI. In addition, we have numerically confirmed scaling laws for the photon energy spread emerging from Compton scattering events, and extended them to include apertures. When applied to numerical Gaussian pulsed photon beams and electron beams with Gaussian spreads, a scaling law for the scattered photon energy spread was verified through a series of numerical computations.  This scaling law has been speculated on previously and has been shown to be valid over a wide range of physically interesting parameters.

A principal motivation for developing this approach is that we could analyze the performance of ODU's compact SRF Compton X-ray source. Front-to-end design simulations have been completed that have been used to make detailed predictions of the photon flux and brilliance expected from the source. Based on the results we have found, SRF-based sources have the potential to produce substantial average brilliance, better than other types of Compton sources, and shown that the brilliance is mainly limited by the beam emittance.

Finally, we have recorded the proper cross sections to apply when the incident or scattered radiation are circularly polarized.

\begin{acknowledgements}
This paper is authored by Jefferson Science Associates, LLC under
U.S.~Department of Energy (DOE) Contract No.~DE-AC05-06OR23177.
Additional support was provided by Department of Energy Office of Nuclear Physics
Award No.~DE-SC004094 and Basic Energy Sciences Award No.~JLAB-BES11-01.
E.~J.~and B.~T.~acknowledge the support of Old Dominion University Office of
Research, Program for Undergraduate Research and Scholarship.
K.~D.~and J.~R.~D.~were supported at ODU by Department of Energy Contract No.~DE-SC00004094.
R.~K.~was supported by the NSF Research Experience for Undergraduates (REU)
at Old Dominion University (Award No.~1359026).
T.~H.~acknowledges the support from the U.S.~Department of Energy, Science
Undergraduate Laboratory Internship (SULI) program.
This research used resources of the National Energy Research
Scientific Center, which is supported by the Office of
Science of the U.S. Department of Energy under
Contract No. DE-AC02-05CH11231
The U.S.~Government retains a non-exclusive, paid-up,
irrevocable, world-wide license to publish or reproduce this manuscript
for U.S.~Government purposes.
\end{acknowledgements}







\bibliography{Krafft_etal_2016}

\end{document}